\documentclass[twocolumn]{aastex631}

\newcommand{\kms}{km~s$^{-1}$}
\newcommand{\msun}{$\rm M_\odot$}
\newcommand{\magarcsec}{mag~arcsec$^{-2}$}

\begin{document}



\title{The Next Generation Virgo Cluster Survey (NGVS). XXXV. First Kinematical Clues of Overly-Massive Dark Matter Halos in Several Ultra-Diffuse Galaxies in the Virgo Cluster}

\author[0000-0001-6443-5570]{Elisa Toloba}
\affiliation{Department of Physics, University of the Pacific, 3601 Pacific Avenue, Stockton, CA 95211, USA}

\author[0000-0002-3790-720X]{Laura V. Sales}
\affiliation{Department of Physics and Astronomy, University of California, Riverside, CA, 92521, USA}

\author[0000-0002-5049-4390]{Sungsoon Lim}
\affiliation{Department of Astronomy, Yonsei University, 50 Yonsei-ro, Seodaemun-gu, Seoul 03722, Republic of Korea}

\author[0000-0002-2073-2781]{Eric W. Peng}
\affiliation{NSF's National Optical-Infrared Astronomy Research Laboratory, 950 North Cherry Avenue, Tucson, AZ, 85719, USA}

\author[0000-0001-8867-4234]{Puragra Guhathakurta}
\affiliation{Department of Astronomy and Astrophysics, University of California, Santa Cruz, UCO/Lick Observatory, 1156 High Street, Santa Cruz, CA 95064, USA}

\author[0000-0002-0363-4266]{Joel Roediger}
\affiliation{Herzberg Astronomy and Astrophysics Research Centre, National Research Council of Canada, 5071 W. Saanich Road, Victoria, BC V9E 2E7, Canada}

\author[0000-0002-3382-9021]{Kaixiang Wang}
\affiliation{Department of Astronomy, Peking University, Beijing 100871, People's Republic of China}
\affiliation{Kavli Institute for Astronomy and Astrophysics, Peking University, Beijing 100871, People's Republic of China}

\author[0000-0002-7089-8616]{J. Christopher Mihos}
\affiliation{Department of Astronomy, Case Western Reserve University, Cleveland, OH 44106, USA}

\author[0000-0003-1184-8114]{Patrick C$\rm{\hat{o}}$t\'e}
\affiliation{Herzberg Astronomy and Astrophysics Research Centre, National Research Council of Canada, 5071 W. Saanich Road, Victoria, BC V9E 2E7, Canada}

\author[0000-0001-9427-3373]{Patrick R. Durrell}
\affiliation{Department of Physics, Astronomy, Geology, and Environmental Sciences, Youngstown State University, Youngstown, OH 44555, USA}

\author[0000-0002-8224-1128]{Laura Ferrarese}
\affiliation{Herzberg Astronomy and Astrophysics Research Centre, National Research Council of Canada, 5071 W. Saanich Road, Victoria, BC V9E 2E7, Canada}








\begin{abstract}

We present Keck/DEIMOS spectroscopy of the first complete sample of ultra-diffuse galaxies (UDGs) in the Virgo cluster. We select all UDGs in Virgo that contain at least 10 globular cluster (GC) candidates and are more than $2.5\sigma$ outliers in scaling relations of size, surface brightness, and luminosity (a total of 10 UDGs). We use the radial velocity of their GC satellites to measure the velocity dispersion of each UDG. We find a mixed bag of galaxies: from one UDG that shows no signs of dark matter, to UDGs that follow the luminosity-dispersion relation of early-type galaxies, to the most extreme examples of heavily dark matter dominated galaxies that break well-known scaling relations such as the luminosity-dispersion or the U-shaped total mass-to-light ratio relations. This is indicative of a number of mechanisms at play forming these peculiar galaxies. Some of them may be the most extended version of dwarf galaxies, while others are so extreme that they seem to populate dark matter halos consistent with that of the Milky-Way or even larger. Even though Milky-Way stars and other GC interlopers contaminating our sample of GCs cannot be fully ruled-out, our assessment of this potential problem and simulations indicate that the probability is low and, if present, unlikely to be enough to explain the extreme dispersions measured. Further confirmation from stellar kinematics studies in these UDGs would be desirable. The lack of such extreme objects in any of the state-of-the-art simulations, opens an exciting avenue of new physics shaping these galaxies.

\end{abstract}



\section{Introduction} \label{sec:intro}


The efficiency with which galaxies transform gas into stars is a strong non-linear function of their halo mass \citep[e.g.][]{ Guo2010,Moster2013,Behroozi2013}. Initially, halos of all masses are expected to contain a baryonic budget consistent with that of the cosmological baryon fraction $\Omega_b \sim 0.17$ \citep{Planck2020_paper6}, meaning that about $17\%$ of their mass is in the form of gas or stars.  However, subsequent internal and external processes associated to galaxy evolution ultimately leads to halos turning a much smaller fraction of those available baryons into stars, bringing the predicted galaxy mass and luminosity function within $\Lambda$CDM in agreement with observations \citep{White1978, Somerville2015, Vogelsberger2020}. 

Halo masses comparable to the one expected for the Milky Way, $\sim 10^{12}$\msun, are believed to be the most efficient of all, condensing about $\sim 20\%$ of the available baryon budget into stellar mass at their centers. For dwarf halos that fraction is much smaller, $1$-$5\%$, being dominated mostly by stellar feedback in the regime of dwarfs and ``classical" dwarf spheroidals and by reionization for the lowest luminosity of all, or ultrafaint dwarfs \citep[$M_* < 10^5$\msun;][]{Bullock2017}. Interestingly, while there are some variations associated to galaxy morphology and environment, the effects are small, with halo mass being still the dominant factor in determining the overall efficiency to transform gas into stars in galaxies \citep[e.g., ][]{Correa2020,Martizzi2020,Engler2021}. 

From this perspective, ultra-diffuse galaxies (UDGs) are powerful probes offering a unique perspective. UDGs are extremely low surface brightness galaxies that have luminosities and stellar masses consistent with those of dwarf galaxies but sizes more typical of massive galaxies \citep[e.g.;][]{Binggeli1985, vanDokkum15}. Possible formation mechanisms for UDGs place them either in dwarf-mass halos, in which case their baryonic efficiency will be in agreement with what is known for dwarf galaxies \citep[e.g.,][]{AmoriscoLoeb2016,DiCintio2017,Carleton2019,Tremmel2020,Wright2021}, or, alternatively, propose that they are ``failed galaxies", meaning that they inhabit more massive halos that were destined to form a brighter galaxy, perhaps comparable to the Milky Way, but their star formation process was truncated at an earlier epoch freezing their stellar content to that of a dwarf \citep[e.g.,][]{vanDokkum15,vanDokkum2016,Janssens22}. The extreme cases of the ``dark-matter free" UDGs, so far only two of these have been discovered: NGC1052-DF2 and NGC1054-DF4, are better explained by a high-velocity collision of gas-rich galaxies \citep{vanDokkum2022}. 

While originally received with enthusiasm, in part propelled by the discovery of several UDGs with a high number of seemingly associated GCs \citep{vanDokkum15,Peng2016}, the hypothesis of massive failed galaxies has lost considerable momentum given more abundant and recent data partially supporting a dwarf-origin for UDGs. This includes diagnostics such as the GC population in field and cluster UDGs being similar to other dwarfs \citep{Beasley2016,Amorisco2018,Lim2018,Somalwar2020,Muller2021,Jones2021,Saifollahi2022,Jones2022},  considerations of passive evolution effects on the central surface brightness of dwarfs in the outskirts of groups \citep{Roman2017a} and their radial distribution \citep{Roman2017b} or the total extent of the light in UDGs \citep{Chamba2020}. Even the revised velocity dispersion and GC number estimates for the iconic DF44 UDG -- one of the best candidates to be a ``failed MW galaxy" -- were found to be smaller than initially estimated and consistent with those of dwarf halos \citep{vanDokkum2019,Saifollahi2021}. 

Such results favoring dwarf-mass halos for UDGs are a confirmation that these galaxies have normal efficiencies at turning baryons into stars, they just do it in a very extended configuration. Several theoretical models have indeed been proposed to explain the large radii of UDGs in dwarf-mass halos, which include internal mechanisms such as halos with large spins \citep{AmoriscoLoeb2016, Rong2017, Benavides2022} or low concentrations \citep{Kong2022} or dwarfs expanded as the result of repetitive and disruptive feedback events \citep{DiCintio2017,Chan2018} or early mergers \citep{Wright2021}; as well as external, environmentally-driven processes, such as tidal heating and stripping \citep{Carleton2019,Safarzadeh2017}, quenching and dimming \citep{Tremmel2020}; or a combination of internal and external mechanisms \citep{Jiang2019, Sales2020}. From the numerical simulations standpoint, it seems like the existence of a population of UDGs living in dwarf-mass halos is not challenging, and instead a natural prediction of $\Lambda$CDM. 

However, observational clues seem to point to the case that at least some UDGs may still require an overly-massive dark matter halo to fully explain their origin. This includes some UDGs with $2$-$4$ times more GCs than dwarfs of similar luminosities \citep{Lim2018,Lim2020,Danieli2022}, even though not quite reaching the levels of a Milky Way-like object. Other cases, such as DGSAT-1, NGC1052-DF2 or NGC 5846-UDG1 show odd distributions of their GC luminosities or colors, or both \citep{Fensch2019,Shen2021,Danieli2022,Janssens22,vanDokkum2022}, hinting also to the need for more exotic origins in these galaxies. The trail of UDGs that might have larger halos than expected is one worth pursuing, as their unusual properties could reveal new physics or mechanisms impacting heavily the efficiency with which halos form stars, a pillar of galaxy formation theories. To date, no cosmological numerical simulation has reported the formation of a ``failed galaxy" UDG living in a massive dark matter halo. 

Arguably, the most direct way to measure dark matter mass in gas-poor galaxies is by means of kinematical studies of their stars or GCs. UDGs with available velocity dispersion estimates are scarce due to their intrinsic low surface brightness (which limits the availability of velocity dispersion for field stars) and finite number of GCs (which negatively impact the accuracy of velocity dispersion estimates). To date, when velocity dispersion has been available, it has supported low dark matter contents consistent with dwarf-mass halos for UDGs \citep{vanDokkum2017,vanDokkum2019,Danieli2019,Chilingarian2019,Collins2020,Gannon2021} with perhaps a handful of exceptions, such as DGSAT-1, a relatively isolated UDG \citep{MartinezDelgado2016,Janssens22} or VLSB-B in the Virgo cluster \citep{Toloba2018}. The need for more observed systems with available kinematical data is indisputable. 

In this paper we present the first systematic study of velocity dispersion in a complete sample of UDGs all located in the Virgo cluster using kinematics of GCs from  Keck/DEIMOS.
Our sample builds upon previous efforts in two main ways. First, we identify UDGs in an objective and unbiased way as those objects that
deviate by more than $2.5\sigma$ from the scaling relations followed by the overall galaxy population.
Second, our sample is {\it complete} in the sense that all known UDGs in the Virgo cluster having a sufficiently large number of GCs have been targeted by our study. In particular, we chose $N_{\rm GC} \geq 10$, following results from cosmological numerical simulations of dwarf galaxies in cluster environments that suggest that $10$ GCs and above provide reliable  dynamical mass estimates \citep{Doppel2021}. In total, our sample includes $10$ UDGs  \citep[of which $3$ were partially presented in][]{Toloba2018}, which more than doubles the existing $7$ UDGs with existing kinematics in the literature which would follow our strict selection criteria.

This paper is organized as follows: We describe our sample selection, data collection, and reduction in Section~\ref{sec:data}. The methods and procedures followed to obtain our direct measurements and our inferred physical quantities are detailed in Section~\ref{sec:analysis}. Main results and comparison with previous measurements of UDGs and normal galaxies in the literature are presented in Section~\ref{sec:results}. In Section~\ref{sec:discussion} we put into a larger context all of our findings, and explore the different possibilities for the formation of UDGs based on our data. We summarize our main findings and conclusions in Section~\ref{sec:summary}.

Throughout this paper, we assume a distance to the Virgo cluster of 16.5~Mpc \citep{Mei2007,Blakeslee2009}, which corresponds to a distance modulus of 31.09~mag. Virial quantities are defined at the radius within which the average density of a halo is $200$ times the critical density of the Universe. 

\section{Observations and Data} \label{sec:data}

\subsection{Sample Selection} \label{sec:sample}

The galaxies analyzed in this work are selected from the complete sample of UDGs identified in the Virgo cluster of galaxies based on the Next Generation Virgo Cluster Survey \cite[NGVS;][]{Ferrarese2012,Lim2020}. The structural parameters of these UDGs are measured using the extremely sharp (median seeing of $0\farcs54$ in the $i'$ band) and deep photometry ($g\sim 25.9$~mag for point-sources and $\mu_g\sim29$~\magarcsec) of the NGVS. The NGVS is a survey done in the $u*g'r'i'z'$ bands with the MegaCam instrument on the Canada-France-Hawaii Telescope (CFHT), and covers from the core to the virial radius of the Virgo cluster, for a total area of 104~deg$^2$. 

Based on the full NGVS galaxy catalogue (Ferrarese et al., in prep.), \citet{Lim2020} identified a total of 44 and 26 UDGs using expansive and restrictive selection criteria, respectively. These UDGs are outliers in scaling relations that utilize the NGVS photometry. The three scaling relations used to define these samples of UDGs correlate luminosity with one of the following three parameters: effective radius ($R_e$, radius that contains half the total luminosity of the galaxy), surface brightness at $R_e$, and mean surface brightness within $R_e$. Outliers are defined as objects that are located beyond the $2.5\sigma$ confidence level from the mean of the scaling relation. \citet{Lim2020} find that 26 objects are outliers in all three scaling relations (the most restrictive sample or primary sample), while 18 more objects are outliers in at least one of the scaling relations (the expansive criteria or secondary sample).

The goal of this work is to estimate the dark matter content in UDGs, and for that we need to measure the velocity dispersion of the galaxies. The extremely low surface brightness of these galaxies (median $<\mu_e>=27.2$~\magarcsec~ in the $g'$ band) makes obtaining stellar dispersions prohibitively expensive in terms of telescope time. However, GCs are compact, gravitationally-bound collections of about a million stars each that are much brighter and significantly faster to target spectroscopically. \citet{Lim2020} characterizes the GC populations of UDGs by identifying point sources in the NGVS $i'$ band (the band with the best seeing, median FWHM$\sim0\farcs54$) based on the ``concentration index", $\Delta i_{4-8}$, defined as the difference between four and eight-pixel diameter aperture-corrected $i$-band magnitudes. Each point source identified in this way is then assigned a likelihood of being a foreground star, a GC in the Virgo cluster, or a background galaxy based on its location in $u*g'i'$ color space \citep[details in][]{Munoz2014,Lim2017,Lim2020}.

In \citet{Toloba2018}, we use simulations based on our Keck/DEIMOS data to assess the statistical significance and biases in recovering the host velocity dispersion when the sample of dynamical tracers (GCs in our case) is small. Our simulations test a range of host $\sigma$ from 10 to 100~\kms, use a minimum of four velocities drawn randomly from the dispersion distribution, and assume the velocity uncertainties are the same as in our data. In these simulations we recover the input velocity dispersion with small biases of $\sim 5$~\kms~ for samples of $<10$~GCs. These data-based simulations agree with recent results from hydrodynamical cosmological simulations  \citep{Doppel2021}. 

We select all UDGs from the full sample by \citet{Lim2020} that contain more than 10 GCs. This selection is done using the total number of GCs after correcting for background contamination, areal coverage, and limiting magnitude \citep[see][and Column (11), N$_{\rm GC,corr}$, in Table~\ref{table:data}]{Lim2020}. The number of Virgo UDGs selected using this criterion is 12. However, we exclude NGVSUDG-A09 (VCC1249) from our sample due to a large contamination in its number of GCs from a nearby massive galaxy, and NGVSUDG-26 (VCC2045) because in our Keck/DEIMOS data we only found two GCs associated to this galaxy. The analysis of the velocity dispersion for NGVSUDG-26 is highly unreliable with such a small sample of GCs.

\begin{figure*}[h!]
\begin{center}
\includegraphics[clip=true,width=0.99\linewidth]{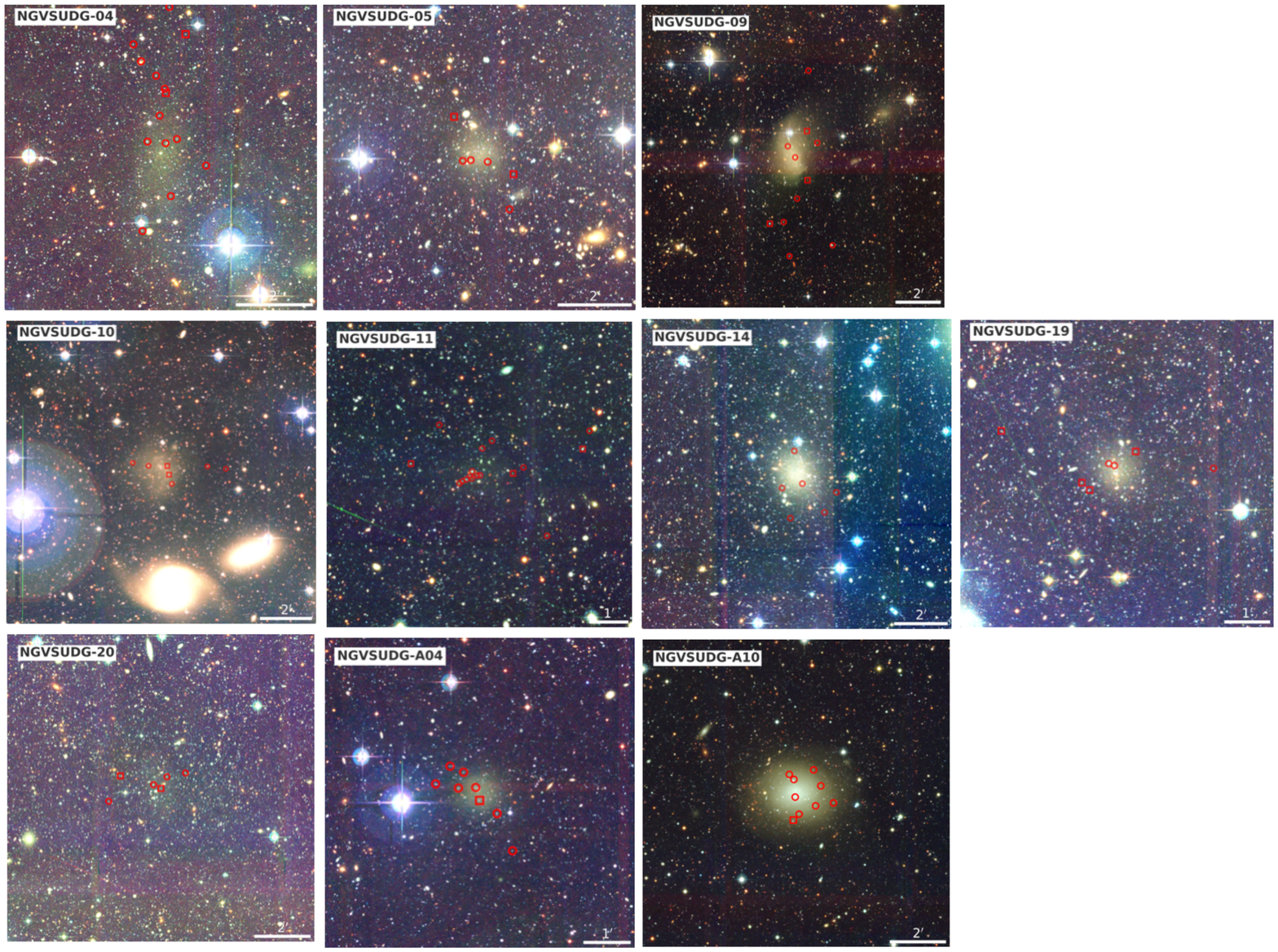}
\caption{Sample of all UDGs in the NGVS footprint with more than 10 GCs photometrically identified as candidate satellites. Each image is a $u^*g'i'$ composite with a field of view of $7.5R_e$, where $R_e$ is the half-light radius of each galaxy. Note that the scale bar is different for each galaxy. To enhance low surface brightness features, we smoothed the images with a Gaussian kernel of 3 pixels. Red symbols show the location of GC satellites. Open circles indicate those objects with high probability of being GC satellites based on their concentration index and $u^*g'i'$ colors. Open squares indicate objects with lower probability of being GC satellites (see Section~\ref{sec:sample} for details). The apparent elongated distribution of GCs is a consequence of using the Keck/DEIMOS spectrograph. The slitlets in the multi-object mask can only be moved perpendicularly to the position angle of the mask to avoid spatial conflicts between the slitlets. The position angle of each mask is selected so that the number of target GCs is maximum. North is up and East is left. None of the UDGs in this sample are nucleated.}
\label{fig:sample}
\end{center}
\end{figure*}

\begin{figure*}[h!]
\begin{center}
\includegraphics[clip=true,width=0.95\linewidth]{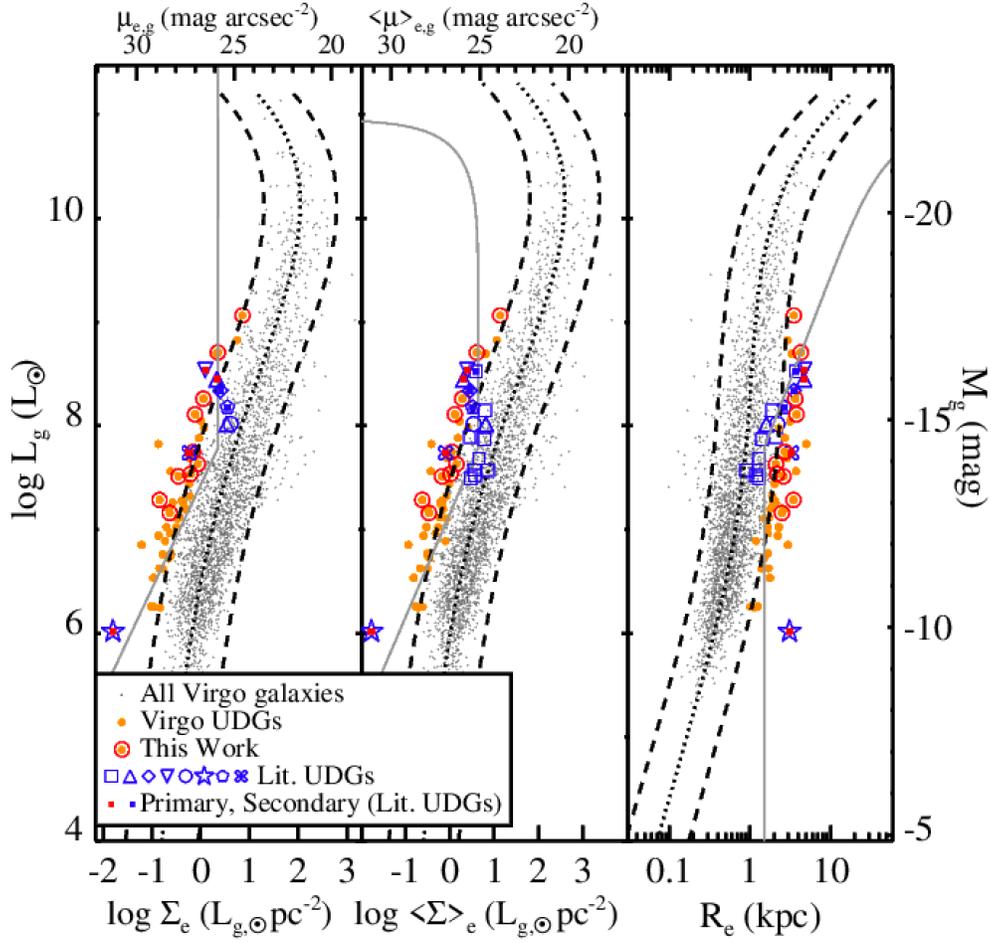}
\caption{Scaling relations in the $g'$ band used to define UDGs. They relate luminosity with surface brightness at $R_e$ ($\Sigma_{e,g'}$), within $R_e$ ($<\Sigma_{e,g'}>$), and the $R_e$. The dotted line shows the best fit and the dashed lines show the $2.5\sigma$ deviations. Galaxies beyond this limit are defined as UDGs by \citet{Lim2020}. If a galaxy is outside all three scaling relations then it is designated as primary and if it is outside one or two only, then it is designated as secondary. The gray lines show the UDG definition adopted by \citet{vanDokkum15}. Note that there is a small region where the definition by \citet{vanDokkum15} intersects the $2.5\sigma$ limit of the scaling relations. Gray dots are Virgo galaxies from the NGVS. Orange dots are the sample of UDGs defined by \citet{Lim2020} while orange dots highlighted with a red circle are the UDGs analyzed here. Blue symbols are objects identified as UDGs in the literature that have a velocity dispersion measurement. References are as follows: squares are from \citet{Chilingarian2019}, triangle is DF44 \citep{vanDokkum2019}, diamond is DFX1 \citep{vanDokkum2017}, upside down triangle is DGSAT I \citep{MartinezDelgado2016,MartinNavarro2019,Janssens22}, circle is NGC5846-UDG1 \citep{Muller2020,Forbes2021}, star is AndXIX \citep{Collins2020}, pentagon is NGC1052-DF2 \citep{Danieli2019,Emsellem2019}, and clover is UDG1137+16 \citep{Gannon2021}. Those with a red or blue square inside adhere to the primary or secondary definitions by \citet{Lim2020}, respectively. We will use only those for comparison to our sample from now on. Note that a large number of galaxies dubbed UDGs in the literature do not follow the definition of \citet{vanDokkum15} or \citet{Lim2020} and are galaxies well within the scaling relations of Virgo galaxies.}
\label{fig:scalingrelations}
\end{center}
\end{figure*}

Our final sample of 10 Virgo UDGs (images shown in Figure~\ref{fig:sample} and their location in the scaling relations in Figure~\ref{fig:scalingrelations}) is the first homogeneous and complete UDG sample where velocity dispersions are obtained from the analysis of GCs. There are no other UDGs in the Virgo cluster for which their dispersion can be estimated using GCs. By necessity, the sample is biased towards those UDGs with relatively large numbers of GCs, which can potentially be a signature of some extreme conditions further discussed throughout the paper. Our sample of UDGs mostly contains UDGs from the primary sample defined in \citet{Lim2020}: eight of them are from the primary sample while only two are from the secondary sample (identified with an A in front of the number in their NGVSUDG name, see Table~\ref{table:data}).

\begin{table*}
\centering
\begin{flushleft}
\caption{Photometric and Structural Properties of the UDGs}\label{table:data}
\begin{tabular}{ccccccccccc}
\hline \hline
Galaxy & Other name & RA & Dec & $M_V$ & $m_{g'}$ & $\mu_{e,g'}$ & $<\mu_{e,g'}>$ & $R_e$ & $R_{1/2}$ & $N_{GC,corr}$ \\
       &            & deg & deg & mag  &  mag     &  mag$/$arcsec$^2$ & mag$/$arcsec$^2$ & arcsec & arcsec &    \\
  (1)  &   (2)      &  (3) & (4) & (5) &  (6)     &  (7)           &   (8)               &  (9)   & (10)   & (11) \\    
\hline
NGVSUDG-04 & VLSB-D & 186.17525 & 13.5168333 & -13.7 & 17.60 & 27.85 & 27.15 & 32.65 & 101.3 & $13.0 \pm 6.9$ \\
NGVSUDG-05 & VCC811 & 186.4067073 & 10.2496165 & -14.3 & 16.96 & 27.10 & 26.60 & 33.96 & 21.0 & $15.8 \pm 8.4$ \\
NGVSUDG-09 & VCC1017 & 186.8814384 & 9.5956422 & -16.7 & 14.54 & 25.83 & 25.19 & 53.69 & 19.3 & $16.5 \pm 11.2$ \\
NGVSUDG-10 & VCC1052 & 186.9800778 & 12.3692963 & -15.2 & 16.06 & 27.00 & 26.43 & 47.44 & 36.2 & $17.9 \pm 11.5$ \\
NGVSUDG-11 & VLSB-B & 187.0419691 & 12.7248463 & -12.3$^\dagger$ & 18.42 & 28.30 & 27.89 & 31.27 & 20.4 & $26.1 \pm 9.9$ \\
NGVSUDG-14 & VCC1287 & 187.6017905 & 13.9818128 & -15.6 & 15.71 & 26.58 & 26.01 & 45.84 & 50.1 & $27.6 \pm 11.1$ \\
NGVSUDG-19 & $-$ & 188.3732917 & 15.2341111 & -13.8 & 17.51 & 27.24 & 26.67 & 27.20 & 21.7 & $16.8 \pm 7.5$ \\
NGVSUDG-20 & $-$ & 188.8035833 & 7.0562222 & -13.2 & 18.09 & 28.82 & 28.26 & 43.48 & 22.1 & $11.3 \pm 8.6$ \\
NGVSUDG-A04 & VCC615 & 185.769125 & 12.0148333 & -14.2$^\dagger$ & 17.25 & 26.86 & 26.34 & 26.33 & 23.3 & $30.3 \pm 9.6$ \\
NGVSUDG-A10 & VCC1448 & 188.1699846 & 12.7711479 & -17.6 & 13.67 & 24.56 & 23.88 & 43.92 & 32.6 & $99.3 \pm 17.6$ \\
\hline
\end{tabular}
\tablecomments{Column 1 shows the name of the galaxy as designated in \citet{Lim2020}. Those galaxies with an A in front of their number are from the secondary sample of UDGs \citep[see Section~\ref{sec:data} and][]{Lim2020}. If the galaxy has another name in the Virgo Cluster Catalog \citep[VCC;][]{Binggeli1985} or was first discovered by \citet{Mihos2015,Mihos2017}, their alternative name is listed in Column 2. Columns 3 and 4 show the right ascension and declination of the galaxy in J2000. Columns 5-9 show the absolute magnitude in the $V$ band, apparent magnitude in the $g'$ band, surface brightness at $R_e$ in the $g'$ band, surface brightness within $R_e$ in the $g'$ band and the half-light radius, all parameters from \citep{Lim2020}. $V$ band magnitudes are obtained from $g'$ magnitudes assuming a color of $g'-V=0.1$~mag. Column 10 shows the radius that contains half the population of GCs (see Section~\ref{sec:masses} for details on its calculation). Column 11 shows the total number of GCs after correction for background contamination, areal coverage, and limiting magnitude as calculated in \citet{Lim2020}. None of the UDGs in this sample are nucleated. However, NGVSUDG-04 and NGVSUDG-A04 both have an overly bright GC-like object that is offset from the center of the stellar light distribution, this object could be considered a nucleus as previously mentioned in \citet{Toloba2018} and \citet{Mihos2022}.\\
$\dagger$ $M_V$ for NGVSUDG-11 and NGVSUDG-A04 are calculated using their updated distance based on our HST data, $12.7^{+1.3}_{-1.1}$~Mpc and $17.7^{+0.6}_{-0.4}$~Mpc, respectively.}
\end{flushleft}
\end{table*}

When comparing to the literature, although the number of UDGs with velocity dispersions available is growing \citep{MartinezDelgado2016,vanDokkum2017,vanDokkum2019,Chilingarian2019,Danieli2019,Emsellem2019,MartinNavarro2019,Collins2020,Muller2020,Forbes2021,Gannon2021}, we will restrict our comparison sample to only those that follow our selection criteria of being outliers in at least one of the luminosity-size-surface brightness scaling relations described above. These include: DGSAT-I \citep{MartinezDelgado2016,MartinNavarro2019}, DFX1 \citep{vanDokkum2017}, DF44 \citep{vanDokkum2019}, NGC1052-DF2 \citep{Danieli2019,Emsellem2019}, UDG7 \citep{Chilingarian2019}, AndXIX \citep{Collins2020}, and UDG1137$+$16 \citep{Gannon2021}.

\subsection{Keck Observations and Data Reduction}

The observations for this project were carried out at the Keck observatory starting in spring 2017. The results for the first three UDGs observed are published in \citet{Toloba2018}. Subsequent observations in spring semesters were affected by unusual events such as an earthquake, the Kilauea volcanic explosion, a snow blackout, and the pandemic. The dates when we gathered data are March 4 2017, February 14 2019, April 24 and 25 2019, February 3, 26 and 27 2020, April 11 2021, and May 10 2021.

All data were collected at the Keck Observatory (Mauna Kea, Hawaii), using the DEIMOS spectrograph \citep{Faber2003} located in the Keck II 10 meter telescope. The instrumental configuration consists of the 600 lines/mm grating centered at 7200~\AA~ with the GG455 blocking filter. We design one slitmask per UDG with slits that are $1''$ wide, with the exception of NGVSUDG-11 for which two slitmasks were designed with different position angles. The wavelength coverage depends on the position of the slit on the slitmask, but roughly, covers $4700 -9200$~\AA~ with a pixel scale of 0.52~\AA/pixel and an average spectral resolution of 2.8~\AA~ (FWHM). The seeing in our observations varies, on average, from 0\farcs5 to 0\farcs9 (FWHM). The exposure time per mask ranges from 4680~s to 26,221~s. This large range is due to on-the-fly adjustments due to seeing variations and sky transparency conditions.

We reduce the data with the SPEC2D pipeline \citep{Cooper2012,Newman13}. In short, this pipeline subtracts the bias and dark current, corrects for flat-fielding, removes cosmic rays, and subtracts the sky. The version of the pipeline we used, described in \citet{Kirby2014,Kirby2015}, includes the following modifications: (1) the wavelength solution is improved by tracing the sky lines across the slit; (2) the object extraction is optimized by tracing the location of the object along the slit accounting for the differential atmospheric refraction. 

The line-of-sight velocities of the GC satellites of NGVSUDG-14 are presented in \citet{Beasley2016}. We take the published measurements of these six GC as they have uncertainties that are comparable to ours (median velocity uncertainty 17~\kms) and  calculate our own velocity dispersion for the UDG using the methods described below. The velocity dispersion obtained in Section~\ref{sec:analysis} is within the error bars of the measurement obtained in \citet{Beasley2016}.


\subsection{Hubble Space Telescope Data} \label{sec:HST}

In \citet{Toloba2018}, we found that NGVSUDG-11 is a UDG with a Milky Way-like line-of-sight velocity (close to 0~\kms). Even though it is not uncommon to find such low, or even negative, radial velocities for objects in the Virgo cluster \citep[e.g.][]{Boselli2006}, this makes it difficult to distinguish GCs belonging to NGVSUDG-11 from Milky Way (MW) stars, as they both appear as point-like sources in ground-based images and can have overlapping colors. We use the exquisite spatial resolution of the \textit{Hubble Space Telescope} (HST) to address the issue, as Virgo GCs will be spatially resolved in HST images. The information obtained from the analysis of this single galaxy is also used in the membership criteria discussed in Section~\ref{sec:membership}.

We use the HST imaging from GO-15417 (PI: E. W. Peng). This program consists of a single HST orbit centered in NGVSUDG-11 and observed with WFC3/UVIS with the F606W filter and a coordinated parallel observation using ACS/WFC also with the F606W filter. Details on the observations, data reduction, and analysis are described in Zhang et al., in prep. We follow the steps of \citet{Jordan2009} where $\sim10,000$ GCs are identified in the Virgo cluster using HST/ACS imaging. In short, we run GALFIT \citep{galfit} to fit PSF-convolved King models \citep{King1966} to determine the sizes of point-like sources. We run extensive simulations to determine how well we can fit GC sizes. The average $R_e$ of a GC in the Virgo cluster is 2.6~pc \citep{Jordan2005}, while all objects with $R_e>1$~pc in the HST imaging are resolved, and, are thus, GCs instead of MW foreground stars.

Zhang et al., in prep. uses the GC luminosity function and the mean sizes of the GCs found in this HST/WFC3 imaging to estimate the distance to NGVSUDG-11 of $12.7^{+1.3}_{-1.1}$~Mpc. This distance puts this galaxy in the foreground of Virgo, which is located at 16.5~Mpc and has a line-of-sight depth of $\sim1$~Mpc \citep{Mei2007}. We use this distance for this galaxy thoughout this paper.

\begin{table*}
\caption{Kinematics, Dynamics, and Other Properties of the UDGs}\label{table:spec}
\begin{flushleft}
\begin{tabular}{ccccccccc}

\hline \hline
Galaxy & $V_{sys}$ & $\sigma$ & $N_{GC,spec}$ & $M_*$ & $M^*_{1/2}$ & $M_{1/2}$ & $M_{1/2}/L_{1/2,V}$ & $f_{DM, 1/2}$ \\
       & km~s$^{-1}$ & km~s$^{-1}$ &          & $\times10^6$~M$_\odot$ & $\times10^6$~M$_\odot$ & $\times10^7$~M$_\odot$ & M$_\odot/$L$_\odot$ & $\%$ \\
  (1)  &  (2)       &  (3)      &   (4)       &  (5)   &  (6)     &  (7)                & (8)      & (9) \\     
\hline
NGVSUDG-04 & $1035^{+6}_{-5}$ & $12^{+6}_{-6}$ & 14 & $58 \pm 12$ & $56 \pm 14$ & $110^{+118}_{-113}$ & $48^{+52}_{-50}$ & 95 \\
NGVSUDG-05 & $982^{+29}_{-29}$ & $64^{+33}_{-19}$ & 6 & $73 \pm 11$ & $18 \pm 4$ & $643^{+937}_{-772}$ & $615^{+896}_{-739}$ & 100 \\
NGVSUDG-09 & $38^{+31}_{-33}$ & $83^{+33}_{-22}$ & 8 & $335 \pm 60$ & $41 \pm 10$ & $996^{+1840}_{-1744}$ & $223^{+412}_{-390}$ & 100 \\
NGVSUDG-10 & $-292^{+6}_{-7}$ & $6^{+11}_{-4}$ & 5 & $208 \pm 38$ & $96 \pm 45$ & $8^{+32}_{-12}$ & $2^{+9}_{-4}$ & 0 \\
NGVSUDG-11 & $40^{+14}_{-14}$ & $45^{+14}_{-10}$ & 14 & $22 \pm 6$ & $9 \pm 6$ & $308^{+196}_{-144}$ & $^\dagger 1875^{+1193}_{-880}$ & 100 \\
NGVSUDG-14 & $1071^{+18}_{-20}$ & $39^{+20}_{-12}$ & 6 & $394 \pm 79$ & $191 \pm 65$ & $559^{+592}_{-359}$ & $75^{+80}_{-48}$ & 97 \\
NGVSUDG-19 & $296^{+37}_{-38}$ & $61^{+47}_{-23}$ & 3 & $62 \pm 14$ & $24 \pm 6$ & $631^{+976}_{-472}$ & $630^{+974}_{-471}$ & 100 \\
NGVSUDG-20 & $946^{+42}_{-41}$ & $89^{+42}_{-27}$ & 6 & $13 \pm 6$ & $2 \pm 1$ & $1318^{+1502}_{-1154}$ & $4698^{+5355}_{-4116}$ & 100 \\
NGVSUDG-A04 & $2089^{+16}_{-15}$ & $36^{+22}_{-18}$ & 8 & $73 \pm 11$ & $31 \pm 6$ & $231^{+282}_{-224}$ & $^\dagger 148^{+181}_{-144}$ & 99 \\
NGVSUDG-A10 & $2310^{+16}_{-17}$ & $48^{+16}_{-11}$ & 9 & $2607 \pm 102$ & $915 \pm 44$ & $548^{+379}_{-248}$ & $18^{+12}_{-8}$ & 83 \\

\hline
\end{tabular}
\tablecomments{Column 1 shows the name of the galaxy as designated in \citet{Lim2020}. Column 2 is the heliocentric systemic velocity. Column 3 is the velocity dispersion of the galaxy. Column 4 is the number of GCs used to calculate $\sigma$. Column 5 is the stellar mass. Column 6 is the stellar mass within the radius that contains half the total population of GCs ($R_{1/2}$). Column 7 is the dynamical mass within the $R_{1/2}$. Note that it is one order of magnitude larger than Columns 5 and 6. Column 8 is the mass-to-light ratio in the $V$ band and within the $R_{1/2}$ for both the dynamical mass and luminosity. Column 9 is the dark matter fraction within the $R_{1/2}$.\\
$\dagger$ $L_{1/2,V}$ for NGVSUDG-11 and NGVSUDG-A04 are calculated using their updated distance based on our HST data.}
\end{flushleft}
\end{table*}

\section{Kinematic Measurements and Stellar Masses} \label{sec:analysis}

\subsection{Line-of-sight Velocity} 

Line-of-sight velocities are measured for all GC candidates using the penalized likelihood software pPXF \citep{Cappellari2004,Cappellari2017}. The templates used in pPXF are a set of 17 high signal-to-noise stellar spectra ($100<S/N<800$~\AA$^{-1}$) observed with the same instrumental configuration as the data, so that both the templates and the science spectra have the same instrumental profile. These 17 stars are selected to cover a large range of spectral types (from B1 to M0) and all luminosity classes (from supergiants to dwarfs). 

Point sources can be miscentered along the slit width direction due to a small misalignment or rotation of the mask. This miscentering is observed as a velocity shift in the atmospheric absorption B and A-bands located at $6850–7020$~\AA~ and $7580–7690$~\AA. We use pPXF to measure the line-of-sight velocity of these absorption features and, if the velocity obtained shows a good clear fit and the velocity is $<|70|$~\kms~ (maximum value allowed due to the $1''$ slit width and instrumental configuration), then the velocity shift is applied prior to measuring the line-of-sight velocity of the science object. The median A-band correction applied is 8.8~\kms.

\begin{figure}[h!]
\begin{center}
\includegraphics[clip=true,width=1.03\linewidth]{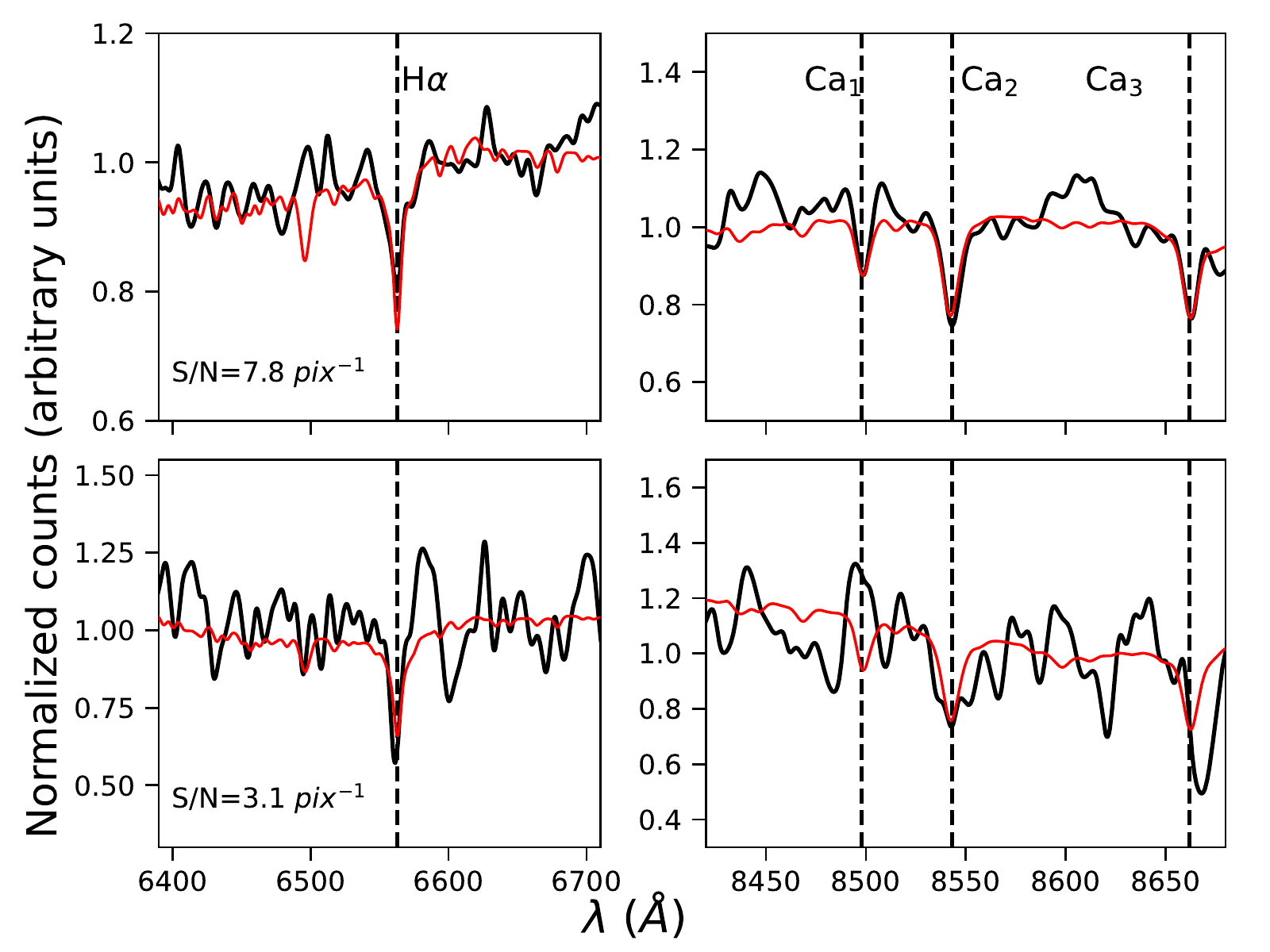}
\caption{Rest frame example spectra of two GCs in our sample. The median S/N of the spectra with reliable velocity measurements is 5.9~pix$^{-1}$. Here we show two randomly picked examples with S/N that are above (upper panel) and below (lower panel) the median S/N of our sample. Our data is shown in black. The best fit model obtained as a combination of the templates observed with the same instrumental configuration is in red. For details on the set of templates used see Section \ref{sec:data}.}
\label{fig:spectra}
\end{center}
\end{figure}

The line-of-sight velocities for all science targets are visually inspected and accepted if at least two clear lines are identified in the fit ($H{\alpha}$ and/or the Calcium triplet lines, see Figure~\ref{fig:spectra}). The $H{\beta}$ and Magnesium triplet regions although observed, are not used for the velocity measurements as the $S/N$ is typically very low in that region. The uncertainties for the velocities are estimated doing Monte Carlo (MC) simulations. In each of the 1000 simulations run per object, the flux of the science target is modified within the flux uncertainty obtained in the reduction process assuming it is Gaussian. In the inspection of the individual fit for each science target, the distribution of the MC simulations is also inspected and used as an additional criterion to accept the velocity. The distribution of the simulations must be single-peaked although it may be skewed. The velocity uncertainty reported are the 16th and 84th percentiles.

\subsection{GC Membership Criteria}\label{sec:membership}

\begin{figure*}[h]
\begin{center}
\includegraphics[clip=true,width=1.05\linewidth]{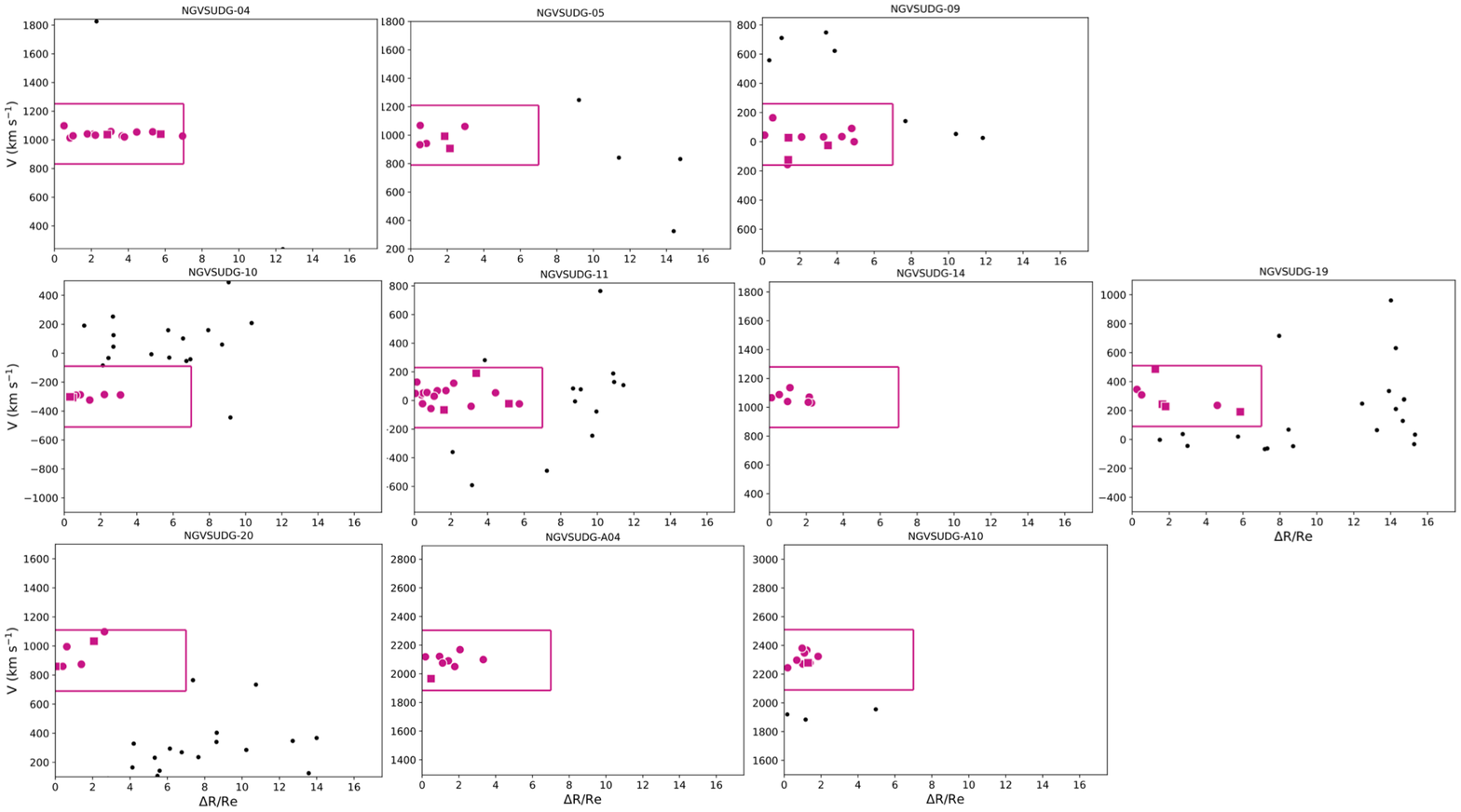}
\caption{Membership Diagram: Objects within the red box are simultaneously close in sky coordinates and radial velocity to the galaxy indicated in the title of each panel, and therefore, are considered its GC satellites. The red box is defined as $\Delta R/R_e \leq 7$ and velocity spread of 210~\kms~ centered on the approximate systemic velocity of the galaxy. The motivation for these numbers are the visual inspection of the radial extension of the GC candidates, specifically trying to include all GCs in NGVSUDG-04 which have very little dispersion in velocity but have a large extension in radius, and the assumption that the velocity dispersion of these galaxies could be as high as $\sigma=70$~\kms, which corresponds to $3\sigma=210$~\kms, based on preliminary results from \citep{Toloba2018}. See Section~\ref{sec:membership} for details. Red dots show GC satellites with high probability of being GCs based on the concentration index and $u^*g'i'$ colors, red squares show lower probabibility of being GCs based on the same criteria. Black dots show other GC candidates in the same mask. Red squares in NGVSUDG-09, NGVSUDG-10, NGVSUDG-11, and NGVSUDG-19 are considered most likely MW contaminants and thus excluded from our analysis (see Setion \ref{sec:membership} for details).}
\label{fig:membership}
\end{center}
\end{figure*}

The criteria used to decide whether a GC is a satellite of a UDG are described in \citet{Toloba2016a,Toloba2018}. The criteria are based on two simultaneous conditions: the GC candidates must be close to the UDG in both sky coordinates and velocity. After careful inspection of the velocity$-$distance space (where distance is measured in terms of the $R_e$ of the UDG), and accounting for the photometric likelihood of the objects to be GCs, we decide to use the conservative numbers of $\Delta R/R_e=7$ (where $\Delta R$ is the distance on the sky of each GC candidate to their potential host and $R_e$ corresponds to that host)  and a velocity spread of 210~\kms~ centered on a tentative systemic velocity for the galaxy under study. The spatial motivation for this choice is that $\Delta R/R_e=7$ includes all GC candidates without large gaps in the spatial direction except for NGVSUDG-19, and it specially includes all GC candidates for NGVSUDG-04. In the case of NGVSUDG-04, all of the GC candidates have such a small dispersion in velocity that it is very likely all of them are satellites. The motivation for the velocity spread of 210~\kms~ is that this value is three times a stellar velocity dispersion of 70~\kms. This relatively large velocity dispersion could be expected for some UDGs in our sample \citep{Toloba2018}. Figure~\ref{fig:membership} shows the membership criteria applied to each UDG. Reducing or enlarging the velocity spread of 210~\kms~ does not change the number of GCs selected as satellites. A maximum of one GC may be added for some galaxies if the velocity spread considered is smaller, but, as described in Section~\ref{sec:MCMC}, removing one GC from our analysis does not change the results.

Whether a photometrically selected GC candidate is a true GC is mostly determined by their radial velocity. 
Background galaxies that appear point-like in our ground-based NGVS images and show red colors in $u^*g'i'$ do not have absorption lines that coincide with the $H{\alpha}$ and Calcium triplet regions, instead, they show prominent emission lines indicating they are sources at higher redshift. However, sources where the photometric likelihood of being a GC (again, based on the concentration index and $u^*g'i'$ colors) is not high and that exhibit radial velocities consistent both with being Milky Way (MW) stars and GCs at the distance of Virgo ($|V|\lesssim 400$~\kms) are difficult to categorize as GC satellites. 

To better assess the nature of these objects (MW stars vs. Virgo GCs) we use HST photometry of NGVSUDG-11 (see Section \ref{sec:HST}), a UDG with a radial velocity close to 0~\kms~\citep{Toloba2018}. In the region covered both by HST/WFC3 and Keck/DEIMOS, we find that two of our spectroscopically measured GC satellite candidates have $R_e<1$~pc, which makes them most likely MW foreground stars. These two objects are the only two with low likelihood of being GCs in our spectroscopic sample in this area. We decide, to be conservative, that all GC candidates with measured radial velocities $|V|\lesssim 400$~\kms~ and a low photometric likelihood of being GCs are most likely MW stars \citep[this velocity range is motivated by the fact that MW stars have radial velocities   $<|300|$~\kms; e.g.][]{Cunningham19}. This leads to excluding from our
analysis all low photometric likelihood objects in the UDGs NGVSUDG-09, NGVSUDG-10, NGVSUDG-11, and NGVSUDG-19 (red squares in Figure~\ref{fig:membership}). For the remaining UDGs, the low photometric likelihood objects cannot be MW contaminants as their velocities are too large to be gravitationally bound to the MW ($|V|\gg 400$~\kms). Further analysis of our HST photometry in NGVSUDG-A04 presented in \citet{Mihos2022} (GO-15258, PI: J. C. Mihos) shows that all objects used in this work for NGVSUDG-A04 are resolved point-like sources, which confirms their GC nature. Table~\ref{table:spec} shows the final number of GC satellites considered for each galaxy.

Point-like sources detected in the ground-based NGVS data with measured radial velocities that are $|V|\gg 400$~\kms~ could be both GC satellites of the target UDG or cluster GC interlopers. The distinction between the two is virtually impossible as both populations have the same sizes, colors, and radial velocities. The likelihood of these objects being GC satellites increases with decreasing distance to the center of the target galaxy. Closer to the center of the galaxy the majority of GCs are expected to be satellites, but regardless, we assess in Section \ref{sec:MCMC} the influence that having GC cluster interlopers would have in our measurements.

\subsection{Systemic Velocity and Velocity Dispersion} \label{sec:MCMC}

We use the GC satellites identified following the criteria described in Section~\ref{sec:membership} as tracers of the gravitational potential of each of the target UDGs. We use a Markov Chain Monte Carlo method \citep[MCMC;][]{MCMC} to determine the systemic velocity ($V_{sys}$) and velocity dispersion ($\sigma$) of each galaxy. We assume that the line-of-sight velocities of the GC satellites come from a Gaussian distribution centered on $V_{sys}$ with a width of $\sigma$. The logarithmic probability of such a distribution is:

\begin{equation}\label{eqn1}
\mathcal{L} (V_{sys}, \sigma)=-\frac{1}{2}\sum_{n=1}^N \log (2\pi (\sigma^2+\delta
v_n^2)) - \sum_{n=1}^N \frac{(v_n-V_{sys})^2}{2(\sigma^2+\delta v_n^2)}
\end{equation}

\noindent where $N$ is the number of GC satellites and $v_n$ and $\delta v_n$ are the radial velocity and error of each GC satellite, respectively. The results obtained can depend on the prior used, in particular for low number of tracers \citep[see][for an exhaustive analysis of the prior effects for $\sigma$ estimations]{Doppel2021}. \citet{Doppel2021} find that a flat prior produces a systematic bias that overestimates the measured $\sigma$, while the Jeffreys prior, due to its net effect of shortening the tails of the posterior distribution function, provides a $\sigma$ that resembles more closely the true value. Here we make our calculations both with a flat prior and a Jeffreys prior using three burns in each case to ensure convergence. The results of the previous burn is used as first guess for the next one. 

The flat prior only puts limits to the expected values of the physical quantities we are calculating. In our case, the flat prior assumes $-500<V_{sys}<3000$~\kms, which is the typical range of radial velocities for Virgo cluster galaxies \citep{Boselli2006}, and $0<\sigma<200$~\kms, which is a plausible range of values for galaxy dispersions. The Jeffreys prior depends on the model used to fit the data, which in our case is a Gaussian function characterized by two parameters, the center and width. The Jeffreys prior for the center of a Gaussian function is 1, and for its width is $1/\sigma$. Equation~\ref{eqn1} works in natural logarithmic space, thus, to introduce the Jeffreys prior, we have to add a factor of $-\log \sigma$. 


\begin{figure*}[h]
\begin{center}
\includegraphics[clip=true,width=0.99\linewidth]{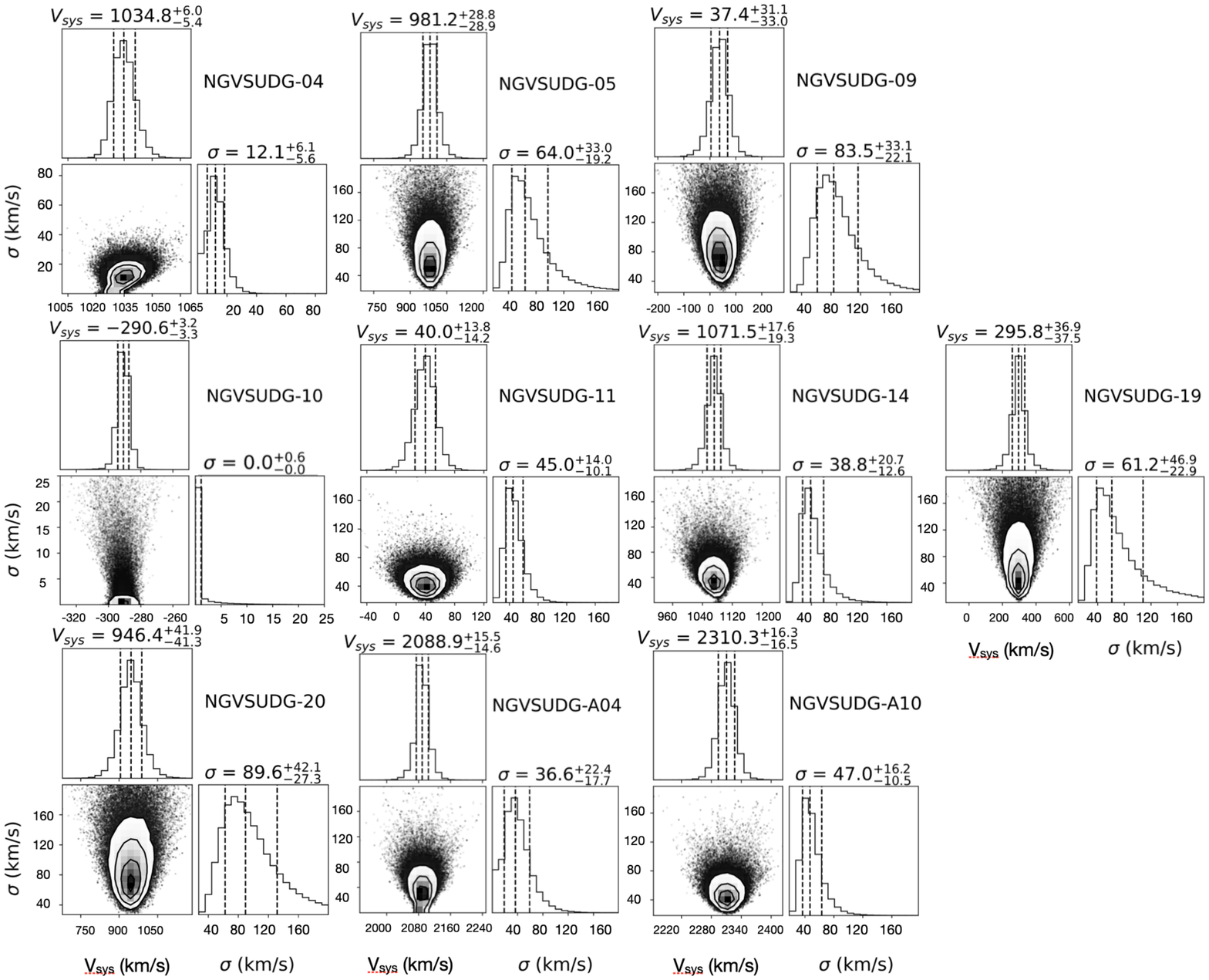}
\caption{Corner plots. Two-dimensional and marginalized posterior probability density function for the systemic velocity $V_{sys}$ and velocity dispersion $\sigma$ estimated using the Jeffreys prior. The kinematic parameters for all galaxies with $V_{sys} > 400$~\kms~ are calculated using the high and low probability GC candidates (red dots and squares in Figure~\ref{fig:membership}). For those galaxies with $|V_{sys}|<400$~\kms~ only the high probability GC candidates are used (only red dots in Figure~\ref{fig:membership}). The velocity dispersion for NGVSUDG-10 cannot be calculated using the Jeffreys prior. We adjust the value obtained using a flat prior and using Equation~\ref{eq:flat_correction} to estimate $\sigma$. See sections~\ref{sec:membership} and \ref{sec:MCMC} for details.}
\label{fig:cornerplot}
\end{center}
\end{figure*}

Figure \ref{fig:cornerplot} shows the two dimensional and marginalized posterior probability density function (PDF) for our sample of UDGs using the Jeffreys prior. For small numbers of GC satellites, the flat prior always finds a larger value for the velocity dispersion than the Jeffreys prior, while $V_{sys}$ is always nearly identical and within the error bars. \citet{Doppel2021} use the Illustris simulations with GCs introduced following the method of \citet{Ramos-Almendares2020} to verify this $\sigma$ overestimation. 

\begin{figure}[h]
\begin{center}
\includegraphics[clip=true,width=0.99\linewidth]{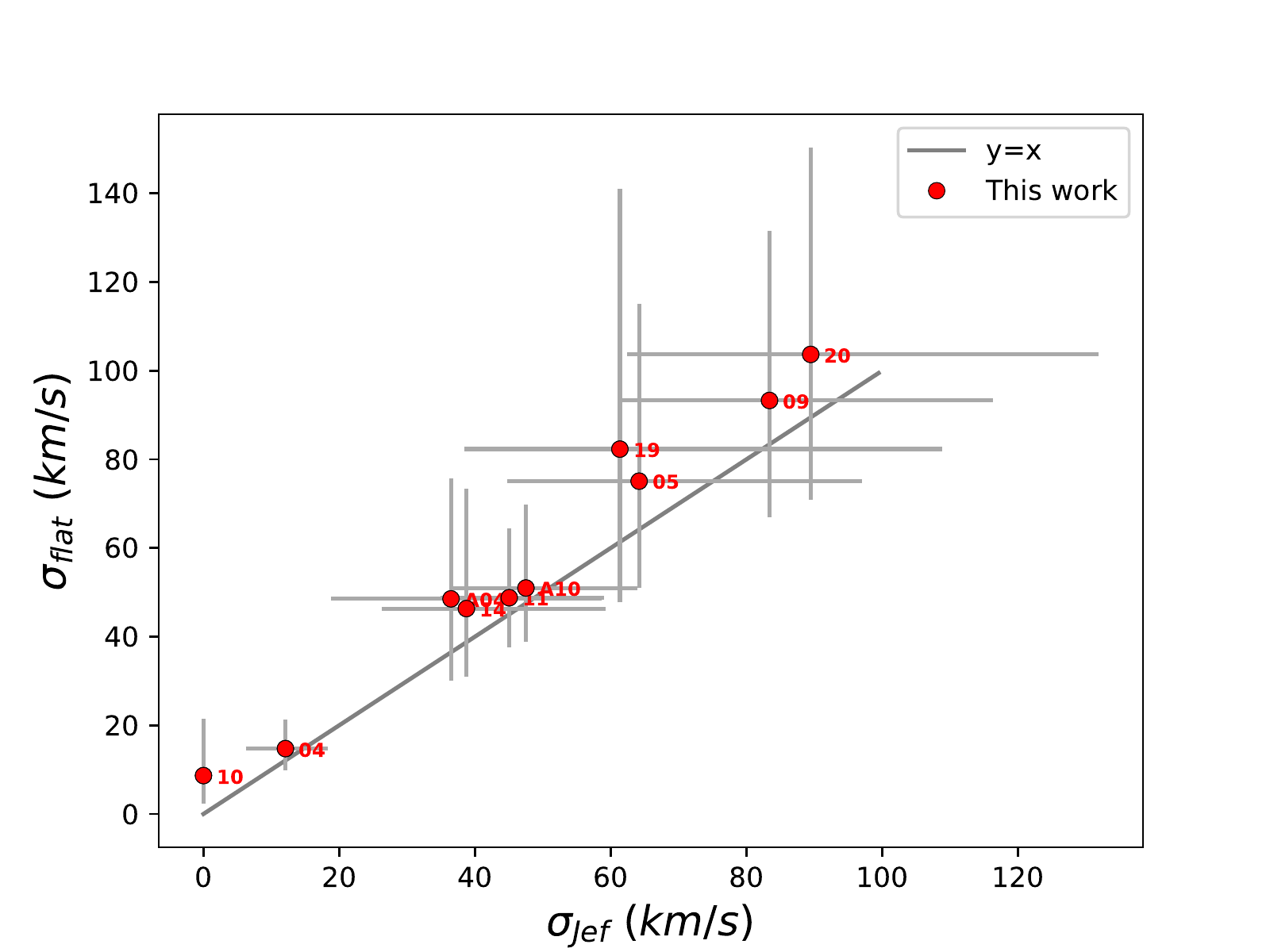}
\caption{Prior comparison. Velocity dispersion obtained using the flat prior vs the Jeffreys prior. The flat prior overestimates $\sigma$ \citep[see also][]{Doppel2021} as highlighted by the fact that all galaxies are above the one-to-one line. A first degree polynomial is fitted to all galaxies but for NGVSUDG-10 (Equation~\ref{eq:flat_correction}). The best fit is used to correct $\sigma$ for NGVSUDG-10, the only galaxy for which the Jeffreys prior gives a dispersion of 0~\kms, most likely because the true value is within the velocity uncertainties.}
\label{fig:flat_jef}
\end{center}
\end{figure}

The velocity dispersion obtained for NGVSUDG-10 is consistent with being 0~\kms~ when the Jeffreys prior is used. This is expected based on the very narrow distribution of velocities shown in Figure~\ref{fig:membership}. When the flat prior is used, the dispersion obtained for this galaxy is  $\sigma_{\rm flat}=8.6^{+12.8}_{6.3}$~\kms. These two low numbers suggest that the real dispersion of these GC satellites are most likely within our error bars and, while $\sigma_{\rm flat}$ slightly overestimates the real value, it is so low that the Jeffreys prior cannot resolve it within the uncertainties. To estimate $\sigma_{\rm jeff}$ for this galaxy we use the $\sigma$ calculated for all other galaxies using both the flat and Jeffreys priors (see Figure~\ref{fig:flat_jef}). We fit a line to the $\sigma_{\rm flat}-\sigma_{\rm Jef}$ relation for all other galaxies and find:

\begin{equation} \label{eq:flat_correction}
\sigma_{\rm flat}=1.14\frac{\sigma_{\rm Jef}}{km~s^{-1}}+2.27 ~km~s^{-1}
\end{equation}

Using $\sigma_{\rm flat}$ for NGVSUDG-10, we estimate its $\sigma_{\rm Jef}$. Table~\ref{table:spec} shows the results for $V_{sys}$ and $\sigma$ for each UDG using Jeffreys prior for all MCMC implementations but for NGVSUDG-10, for which we use the Jeffreys dispersion calculated using Equation \ref{eq:flat_correction}.

As discussed in Section~\ref{sec:membership}, the samples used here are fully cleaned from background galaxies, however, there is still a remote possibility that we have cluster GC interlopers, or MW stars in the case of NGVSUDG-09, NGVSUDG-10, NGVSUDG-11 and NGVUDG-19. We assess the contribution of these possible contaminants by removing one GC candidate for each galaxy at a time and running the same MCMC method. The results obtained in all cases and for all galaxies are always within the errorbars. If these contaminants are present in any of our galaxies, they are not significantly affecting the results. See Section~\ref{sec:discussion_extreme} for a detailed discussion on the possible contaminants.

 In \citet{Toloba2018} we investigate the possibility of the GCs in some UDGs showing rotation. In that paper, we find hints of some coherent disturbance in the velocities of the GCs of NGVSUDG-04 along the semimajor axis. We analyze this possibility in Appendix~\ref{A1} for all 10 galaxies in our sample, however, we only find some weak hints of velocity disturbances in  NGVSUDG-04 and NGVSUDG-10. In both cases, the velocity gradient estimated has large uncertainties and is consistent with zero. Curiously, these are the two UDGs in our sample with, somewhat, elongated structures that may be indicative of tidal interactions. Getting a result for rotation consistent with zero or not a convergence of the MCMC method suggests that the rotation, if present, cannot be measured with these data, which puts an upper limit for the rotation of GCs to $\sim 12$~\kms, which is the median radial velocity uncertainties for our sample. See Appendix~\ref{A1} for more details on this analysis.

\subsection{Stellar Mass Estimates}

Stellar masses are determined for all entries in the NGVS galaxy catalogue via modelling of their $u^*g'i'z'$ spectral energy distributions (SEDs, Roediger et al., in prep.). The SEDs are measured following one of two approaches: fitting elliptical isophotes with a bespoke code based on IRAF/ELLIPSE \citep{ellipse} or fitting 2D S\'ersic models with GALFIT \citep{galfit}.  Readers interested in the modelling of the light distributions of NGVS galaxies are referred to \citet{Ferrarese2020}.  For galaxies with $B_{VCC} < 16$~mag \citep[the $B$ band magnitude from the Virgo Cluster Catalog, VCC;][]{Binggeli1985} we draw on the growth curves from IRAF/ELLIPSE, and measure fluxes integrated within the first isophote where the $g'$-band growth curve flattens.  For all other galaxies we use the total fluxes (integrated to infinity) of the best-fit model found by GALFIT.  Errors in the integrated fluxes are estimated by summing the per-pixel contributions from Poisson noise of the source and sky, and read noise, while enforcing lower limits equal to the precision of the NGVS photometric calibration. 

We model the SEDs using the code {\tt PROSPECTOR} \citep{prospector}.  This code, based on the Flexible Stellar Population Synthesis (FSPS) model suite \citep{Conroy2009}, allows its users to generate photometric and spectroscopic data for synthetic stellar populations for a variety of star formation histories (SFHs), metallicities, initial mass functions (IMFs), dust attentuation curves, etc. and to measure the Bayesian PDF of model parameters via MCMC or dynamic nested sampling algorithms.  We employ FSPS simple stellar population spectra based on the MIST isochrones \citep{MIST_0,MIST_I}, the MILES stellar library \citep{MILES_I,MILES_II}, and the Chabrier initial mass function \citep[IMF;][]{ChabrierIMF}, and model the NGVS photometry using four free parameters: one for stellar mass, two for our assumed delayed-$\tau$ SFH (age, timescale), and one for metallicity.  We adopt flat priors for all parameters, either on linear (age, metallicity) or logarithmic (mass, timescale) ranges.  We sample the posteriors using the MCMC algorithm offered by PROSPECTOR and report stellar masses as the median values of the marginalized PDFs.  Adding an additional free parameter to our modelling to represent attenuation by a simple dust screen changes our stellar mass inference by less than $25\%$ for $80\%$ of the NGVS galaxy sample.  It should also be noted that fits including dust attenuation do not produce a statistically significant improvement in maximum likelihood for $>95\%$ of the sample (Roediger et al., in prep.).

\subsubsection{Stellar Mass Within $R_{1/2}$} \label{sec:Mstarhalf}

When comparing physical magnitudes it is important to measure all of them within the same aperture. While the literature uses stars to estimate the velocity dispersion of galaxies, we use GCs. When stars are used, the aperture that contains half the light is the $R_e$ and the stellar mass within $R_e$ is then, by definition, half the total stellar mass estimated ($M^*/2$). However, when GCs are the dynamical tracers, the mass within the aperture that contains half the tracers ($M^*_{1/2}$) has to be calculated. We estimate $M^*_{1/2}$ using the curve of growth in the reddest band for which we have NGVS photometry, the $z$ band, and integrate it within a radius $R_{1/2}$ that contains half the tracers ($R_{1/2}$ calculations are described in Section~\ref{sec:masses}). We follow the Equations described in \citet{Graham05}.


\section{Results}\label{sec:results}

\begin{figure*}[h]
\begin{center}
\includegraphics[clip=true,width=0.99\linewidth]{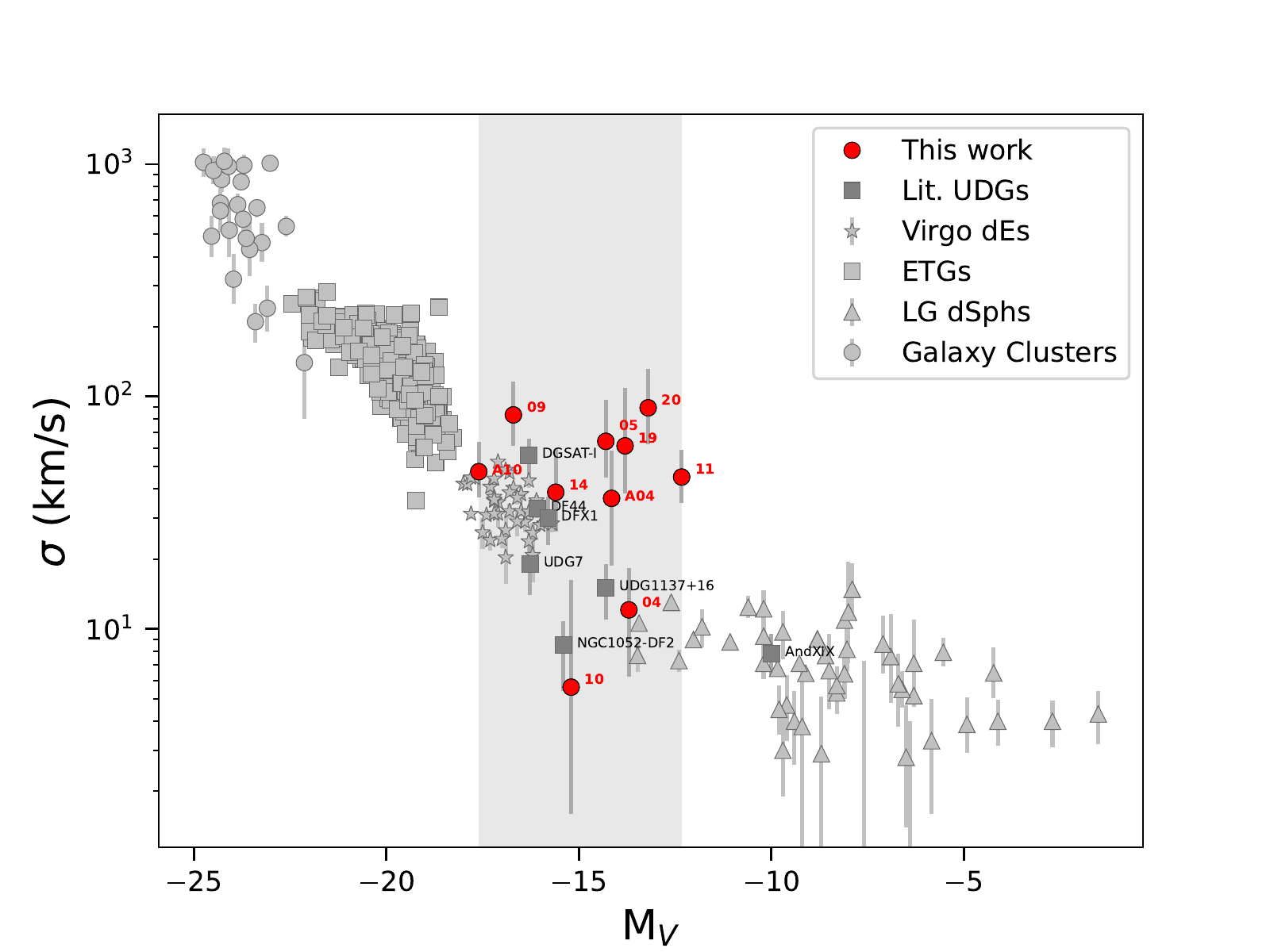}
\caption{Velocity dispersion-luminosity in the $V$ band relation. The UDGs analyzed in this work are shown in red (only the number part of their name is shown for short) in comparison to galaxy clusters \citep{Zaritsky2006}, local volume early type galaxies \citep{Cappellari2013}, Virgo cluster dwarf early-types \citep{Toloba2014b}, Local Group dwarf spheroidals \citep{Collins2013,Wheeler2017}, and UDGs from the literature \citep[none of them in the Virgo cluster;][]{MartinezDelgado2016,vanDokkum2017,vanDokkum2019,Danieli2019,Chilingarian2019,Collins2020,Gannon2021}. See text for details. The gray region highlights the luminosity range of our sample of UDGs. While the UDGs studied in the literature and $50\%$ of the sample analyzed here are consistent with the trend of other quenched galaxies, the remaining $50\%$ of our sample are clearly offset from this relation. These exhibit a velocity dispersion that is too large for their luminosity.}
\label{fig:sig_MV}
\end{center}
\end{figure*}

The velocity dispersion of non-rotating systems is an important proxy for mass. Quiescent galaxies show a well defined luminosity-velocity dispersion relation, which evolves from $\sigma \sim 1000$ km/s in bright galaxies with $M_V \sim -25$ at the centers of galaxy clusters to $\sigma \sim 5$-$10$ km/s in dSph in the Local Group with $M_V > -10$. Figure~\ref{fig:sig_MV} shows a compilation of data from the literature showing this trend, including clusters of galaxies \citep{Zaritsky2006}, local volume early type galaxies \citep[ETGs from the ATLAS$^{\rm 3D}$ survey;][]{Cappellari2013}, Virgo cluster dwarf early type galaxies \citep[dEs;][]{Toloba2014b} and Local Group dwarf spheroidal galaxies \citep[LG dSphs;][]{Wolf2010,Wheeler2017}.

We overlay in Figure~\ref{fig:sig_MV} the velocity dispersion as a function of $V$-band magnitude for our sample of UDGs (red symbols) with the caveat that our $\sigma$ values are computed using GCs and not stars, unlike the compilation of data from literature for all early-type galaxies. 
We find a very large intrinsic scatter in the $\sigma$ observed for our UDG sample, which is substantially larger than the variations found in dEs or dSphs at similar luminosities. For comparison, we also highlight using dark-grey squares previous measurements for UDGs in the literature that satisfy our definition as scaling relation outliers (see Section~\ref{sec:data} for details). The $\sigma$ of these UDGs from the literature are all obtained from stars. None of these UDGs are in the Virgo cluster. They are either in isolation or a satellite of a more massive companion (M31, NGC1052, or UDG1137$+$16) with the exception of UDG7, which is in the Coma cluster. There are GC counts only for three out of the seven UDGs from the literature, and all three have large numbers of GCs.  Notice that all literature UDGs with measured kinematics follow nicely the sequence of ``normal" quenched galaxies, in agreement with the interpretation that such UDGs are the most extended tail of the ``normal dwarfs" distribution. 

Our Virgo UDG sample, however, shows more extreme behavior. While about $3$-$4$ UDGs outline the normal galaxy sequence (NGVSUDG-A10, NGVSUDG-14, NGVSUDG-04, with NGVSUDG-10 slightly below due to suspected tidal stripping), about half of our sample is signficantly off the relation, showing very large velocity dispersion given their luminosity. These outliers include: NGVSUDG-05, NGVSUDG-11, NGVSUDG-19, NGVSUDG-20 and NGVSUDG-A04, with NGVSUDG-09 being borderline compatible. The high velocity dispersion in these UDGs suggests a larger dark matter mass content within $R_{1/2}$ and is at face value consistent with having overly-massive halos given their stellar content, in agreement with some of the first hypothesis for the formation of UDGs as ``failed MW galaxies" \citep{vanDokkum15,Peng2016}. Noteworthy, this is the first {\it kinematical} signature of such scenario ever reported \citep[note that originally DF44 was believed to have a large velocity dispersion, but that has been revised,][]{vanDokkum2019}.

Besides the UDGs with large velocity dispersions, it is also interesting to highlight the broad range of kinematics measured in our sample. To attest to this kinematical diversity in UDGs, it is worth noting that NGVSUDG-10 in our sample shows a dispersion and luminosity that are very similar to NGC1052-DF2, which has been claimed to be devoid of dark matter \citep[see][]{vanDokkum2018a,vanDokkum2018b,Danieli2019,Emsellem2019}, showing the exact opposite behavior to the subsample highlighted above: a velocity dispersion too low for their stellar content. As discussed in \citet{Lim2020}, NGVSUDG-10 shows a peculiar morphology with possible spiral arms or tidal streams with some star formation, this suggests that tidal interactions with other galaxies may be a viable mechanism for the formation of UDGs with low velocity dispersion \citep{Sales2020, Doppel2021, Moreno2022}.

\subsection{The Dynamical Mass and Dark Matter Content of UDGs}\label{sec:masses}

We calculate the total (dynamical) mass, using the following formula:

\begin{equation}\label{eq:mass}
M_{1/2} = 930 \frac{\sigma^2}{km^2 s^{-2}} \frac{R_{1/2}}{pc}~M_{\odot}
\end{equation}

\noindent
where $\sigma$ is the velocity dispersion obtained from the MCMC analysis described in Section~\ref{sec:MCMC} and $R_{1/2}$ is the radius that contains half the dynamical tracers \citep[dynamical mass estimation from][]{Wolf2010}. This mass estimator assumes the dynamical tracers have spherical symmetry and are in dynamical equilibrium. It is mostly insensitive to the anisotropy parameter of the orbits of the tracers and to projection effects \citep[see][]{Wolf2010}. This formalism was shown in \citet{Doppel2021} to be able to accurately recover the true dynamical masses in dwarfs with $\sim 10$ or more GCs.

The $R_{1/2}$ parameters in our analysis corresponds to the radius that contains half the population of GCs. This radius is calculated by fitting a S\'ersic profile with index $n=1$ \citep{Sersic63} assuming circular distribution of only high probability candidate GCs (see Section \ref{sec:sample}). The $R_{1/2}$ parameter values for NGVSUDG-10 and NGVSUDG-19 are extremely uncertain; we estimate the $R_{1/2}$ value for these galaxies based on their half-light radius ($R_e$). We fit a line to the $R_e-R_{1/2}$ relation for all UDGs excluding, of course NGVSUDG-10 and NGVSUDG-19, and also NGVSUDG-04 as it is the only galaxy that is more than $3\sigma$ away from the linear fit. The best fit is the following:

\begin{equation}\label{eq:Re-Rh}
R_{1/2} = 0.67\frac{R_e}{arcsec} +4.22 ~arcsec
\end{equation}

With this equation, we obtain the $R_{1/2}$ for NGVSUDG-10 and NGVSUDG-19 listed in Table~\ref{table:data}. If we assume that, instead of these two galaxies following this relation (Equation~\ref{eq:Re-Rh}), their $R_{1/2}$ is the same as $R_e$ \citep[as it is common for early-type galaxies, e.g.;][]{Peng2006} or even larger, then $M_{1/2}$ would be larger, too, making them even more dark matter dominated.

\begin{figure*}[h]
\begin{center}
\includegraphics[clip=true,width=0.49\linewidth]{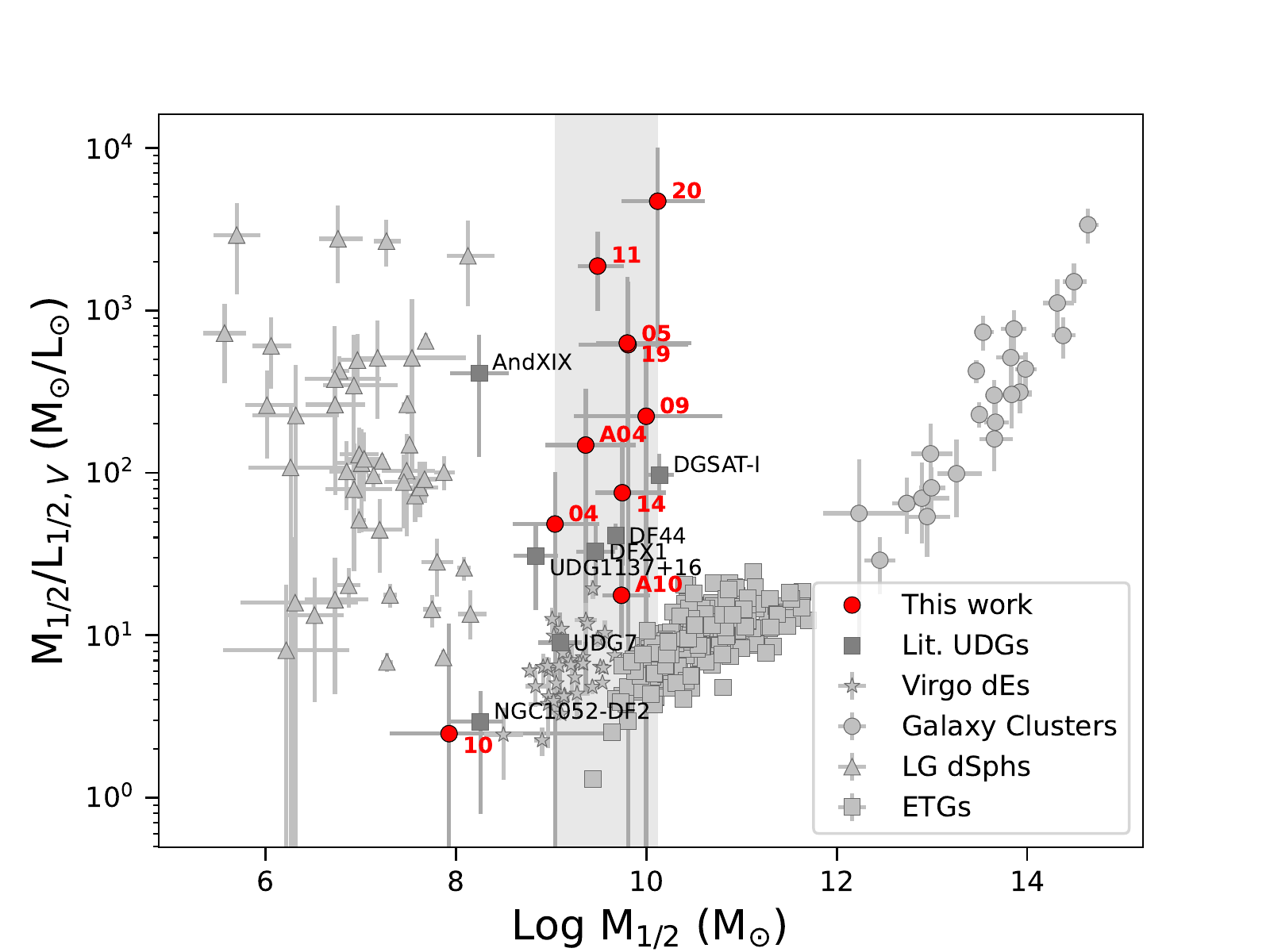}
\includegraphics[clip=true,width=0.49\linewidth]{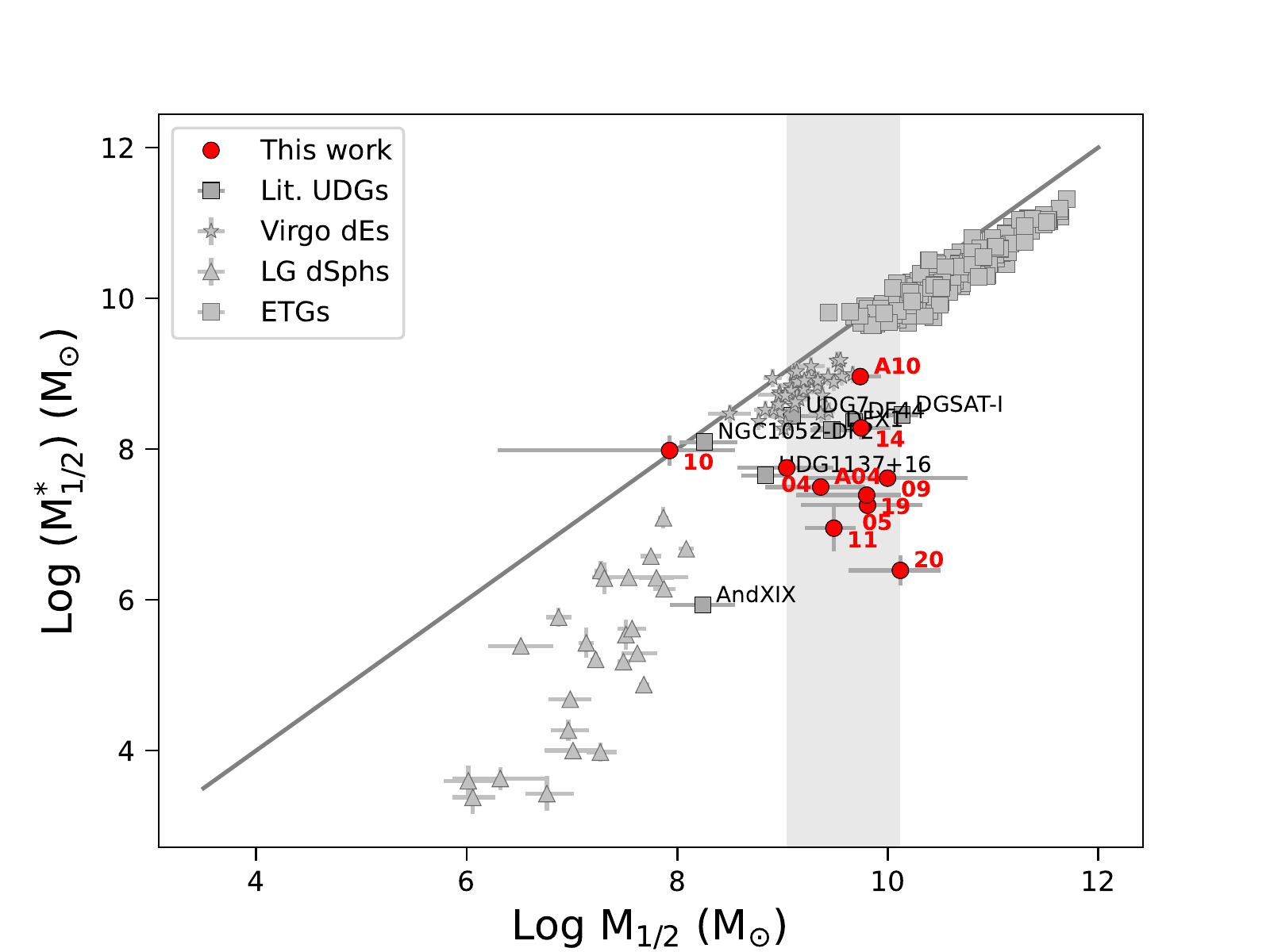}
\caption{Total mass-stellar mass relation. Left panel: total (aka dynamical) mass-to-light radius as a function of the total mass in the $V$ band. Note several of our UDGs have mass-to-light ratios $> 100$, comparable to dSphs but in significantly more massive halos, as judged by their high $M_{1/2}$. Right panel: stellar contribution to the dynamical mass within $R_{1/2}$. For most galaxies $M^*_{1/2}$ is simply $M^*/2$, but for UDGs we have calculated the stellar mass within the half-number radius of the GCs, which is different than the half-light radii (see Section~\ref{sec:Mstarhalf} and Table~\ref{table:data} for details). The gray line shows the $\log M^*_{1/2} = \log(M_{1/2})$ relation. Symbols as in Figure~\ref{fig:sig_MV}. The gray shaded areas highlight the $M_{1/2}$ range for all the UDGs in our sample for which the dynamical mass is not explained by stellar mass alone, i.e. they contain significant amounts of dark matter. $M_{1/2}$, $M^*_{1/2}$, and $L_{1/2,V}$ in both panels are calculated within the radius that contains half the dynamical tracers (GCs for our galaxies highlighted in red and stars for all other systems in gray). Galaxies near the 1:1 line are consistent with all or a high fraction of the total mass contributed by the stars. UDGs scatter downwards away from this line, suggesting a dark matter fraction within $R_{1/2}$ as high as those in dSphs in the Local Group.}
\label{fig:Mdyn-Mstar}
\end{center}
\end{figure*}

The left panel in Figure~\ref{fig:Mdyn-Mstar} shows the relationship between the total dynamical mass ($M_{1/2}$) and the mass-to-light ratio for our Virgo UDG sample (red symbols), UDGs in the literature (dark gray) and for the full range of quiescent objects compiled from the literature (light gray). In this space, while most UDGs, including those in the literature, appear to be more dark matter dominated at a given $M_{1/2}$, some of the Virgo UDGs in this work show more extreme mass-to-light ratios $\geq 100$-$1000$, so high that are on par with some of the faintest dSph in the Local Group or with the values found for clusters of galaxies as a whole.  

Indeed, while mass-to-light ratio as a function of dynamical mass is known to show the ``U"-shape behaviour outlined by the compilation of normal quenched galaxies, UDGs appear as objects with intermediate $M_{1/2}$ but high mass-to-light values, breaking the U-pattern known before. On the other hand, UDGs seem to preserve the minimum $M_{1/2}/L_{1/2,V} \sim 2$ observed for other spheroidal galaxies, with not even the seemingly dark-matter poor NGVSUDG-10 or NGC1052-DF2 breaking such a trend.

A different way to look at the mass to light ratio of UDGs compared to other galaxies is to examine the stellar contribution to the dynamical mass estimated within $R_{1/2}$. This is shown in the right panel of Figure~\ref{fig:Mdyn-Mstar}, where $M^*_{1/2}$ is the stellar mass within the half-tracer radii.  The gray line shows a one-to-one relation, which is the limit indicating that a galaxy does not have dark matter. As expected, NGVSUDG-10 and NGC1052-DF2 lie on the one-to-one relation, followed closely by massive (squares) and dwarf (stars) early-type galaxies (ETGs and dEs, respectively), where most of the dynamical mass can be explained by the stars, at least within their $R_e$  \citep[][]{Cappellari2006,Cappellari2013,Toloba2014b}. 

UDGs in our sample span a wide range of stellar mass contribution at a relatively narrow dynamical mass range. Two of our Virgo UDGs and most of the literature UDGs show also a significant contribution of the stars to their dynamical mass within $R_{1/2}$, making them comparable to dEs, but more extended. However, the remaining six UDGs in our sample (NGVSUDG-A04, NGVSUDG-05, NGVSUDG-11, NGVSUDG-19, NGVSUDG-20, and NGVSUDG-09)  are clearly off this relation and show a much larger dark matter fraction within $R_{1/2}$ than dEs and ETGs. Actually, these six UDGs, which are the same ones highlighted as outliers in Fig.~\ref{fig:sig_MV}, seem to be as far away from the no-dark matter line as the Local Group dSphs (triangle symbols). 

We estimate the dark matter fraction within the radius that contains half of the dynamical tracers in each galaxy using the relation:

\begin{equation}\label{eq:fDM}
f_{DM, 1/2} = \frac{M_{1/2}-M^*_{1/2}}{M_{1/2}}
\label{eq:f_dm}
\end{equation}

Table~\ref{table:spec} shows the dark matter fraction for each of our UDGs, which ranges from consistent with $0\%$ dark matter in NGVSUDG-10 to $100\%$ for NGVSUDG-05, NGVSUDG-09, NGVSUDG-11, NGVSUDG-19, and NGVSUDG-20, with a median dark matter fraction for all the UDGs in our sample  $\bar{f_{DM,1/2}} \sim 98\%$. For comparison, the median dark matter fraction for ETGs and dEs are $47\%$ and $46\%$, respectively, suggesting similar contributions from baryons and dark matter within $R_{e}$, while the dark matter content for the Local Group dSphs is $100\%$, consistent with the picture of dark matter dominated objects. Most of our UDGs and in particular the $6$ high-$\sigma$ candidates highlighted before, have high dark matter fractions, 
$f_{DM,1/2} \sim 100\%$, which are comparable to those only seen in the less-massive dSphs in the Local Group. 

\begin{figure*}[h]
\begin{center}
\includegraphics[clip=true,width=0.49\linewidth]{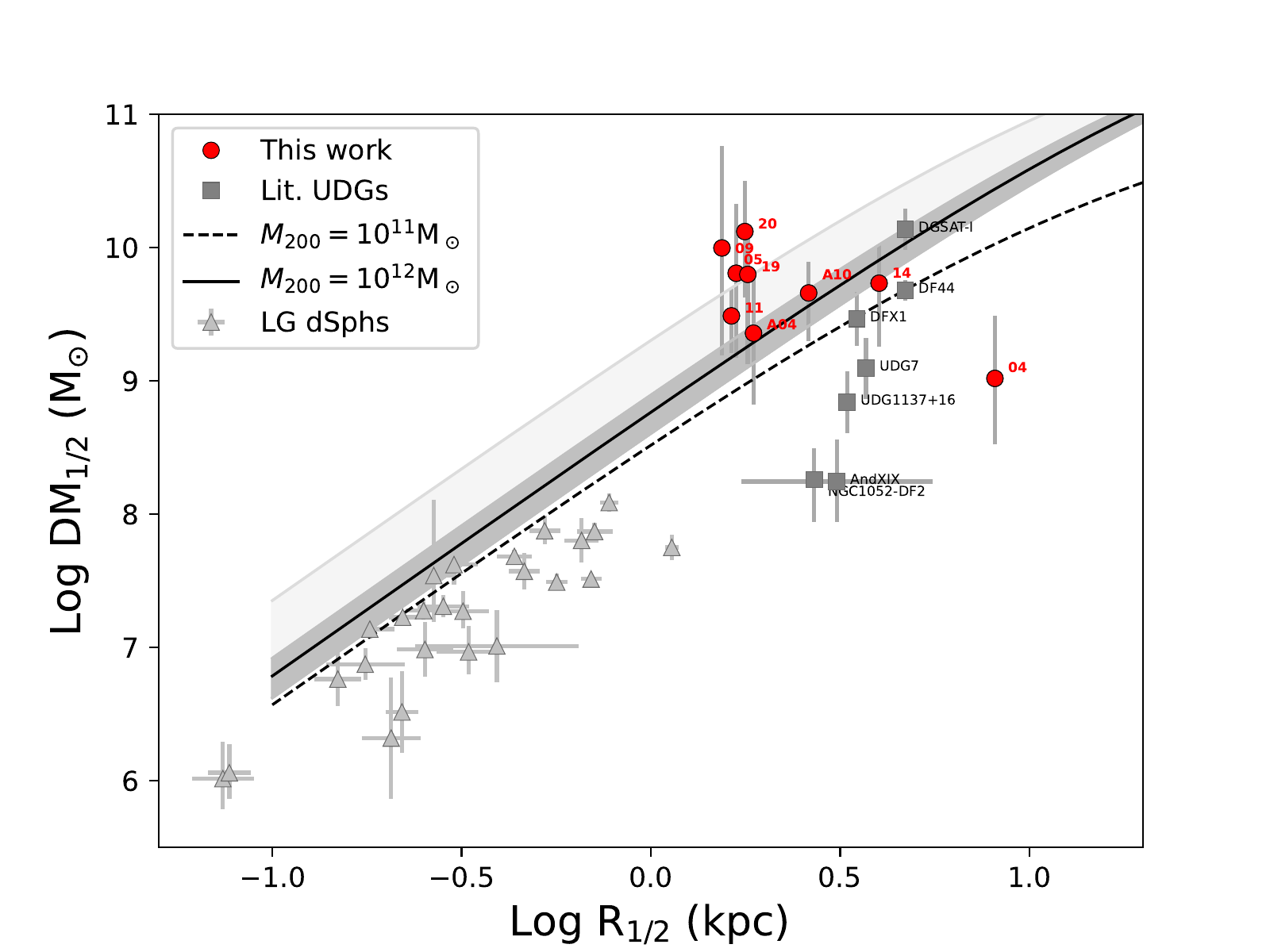}
\includegraphics[clip=true,width=0.49\linewidth]{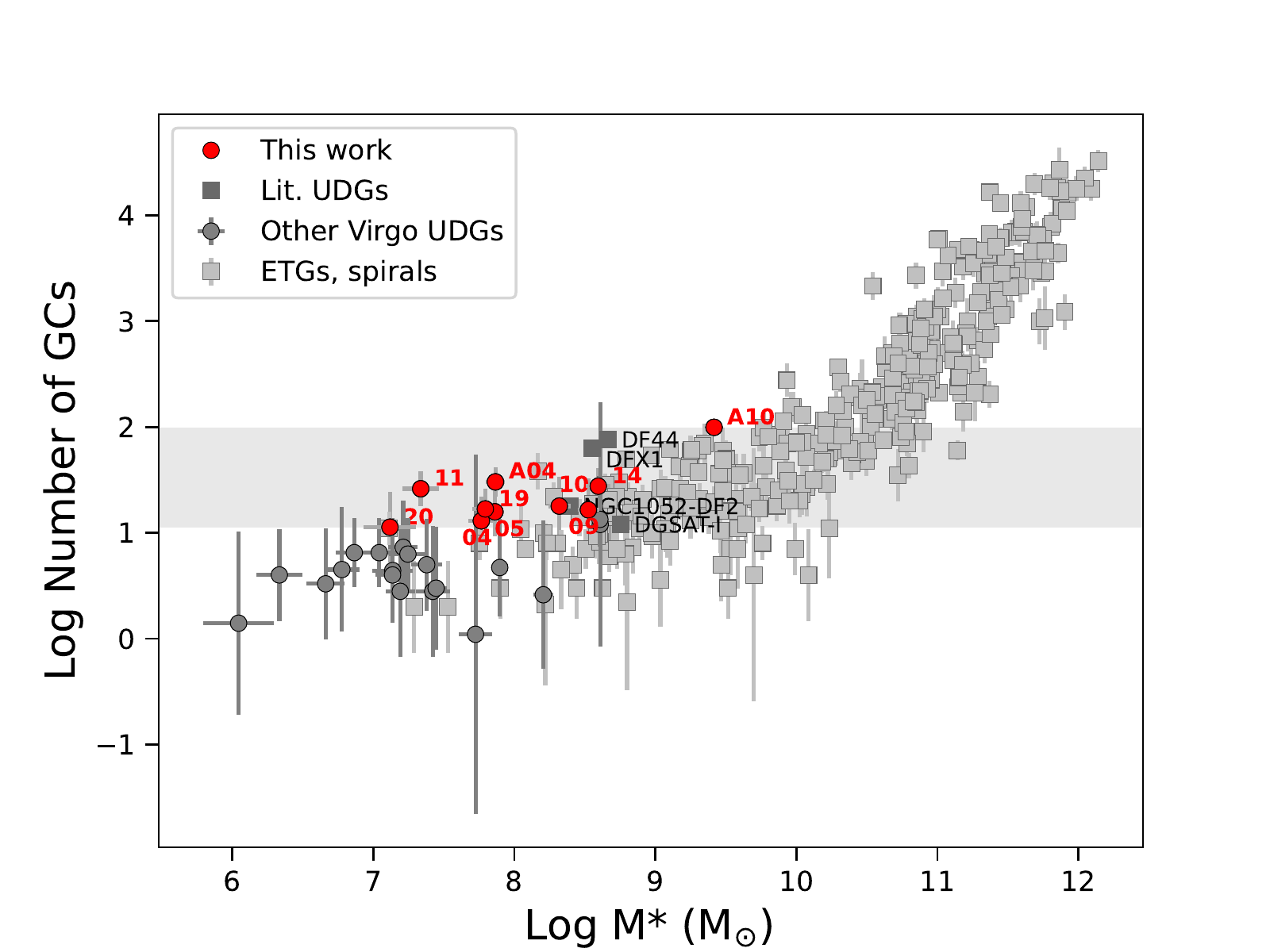}
\caption{{\it Left:} dark matter mass inferred at $R_{1/2}$ after subtracting the stellar contribution at $R_{1/2}$ from the dynamical mass for  each galaxy. UDGs in our sample are shown in red symbols, literature UDGs in dark gray symbols and for reference dSphs in the local group are shown in light gray squares. For comparison, the dashed and solid curves show NFW halos with virial mass $M_{200}=10^{11}$ (concentration $c=10$) and $10^{12}\; \rm M_\odot$ ($c=8.3$), respectively \citep{Dutton2014}, with variations of concentration in the more massive halo $c \sim 6$ - $10$ (dark gray) and the more extreme $c=20$ (light gray). Several of our Virgo UDGs require large mass halos and high concentration to explain their large dark matter content. On the other hand, all but one literature UDG are consistent with dwarf-mass halos (with the exception of DGSAT-1). Also clear in this plot, dSphs while showing similar dark matter fractions than our most extreme UDGs, they populate less massive halos and their stars trace much smaller regions than that traced by the stars or GCs in the UDGs. {\it Right:} GC number as a function of stellar mass for a sample of early type galaxies. Virgo UDGs in this work have biased-high GC contents which span $\sim 10$-$50$ GCs except for our most massive UDG with inferred $\sim 100$ GCs. The gray shaded area highlights the range in the number of GCs for our sample of UDGs.}

\label{fig:mass-NGC}
\end{center}
\end{figure*}

What halos would accommodate the measured kinematics in UDGs? Fig.~\ref{fig:mass-NGC} shows a compilation of dynamical mass estimates at the half-tracer radius for dSph (light triangles) and literature UDGs (dark gray squares) along with our new UDG sample (red symbols). For comparison, we include the mass profile expected for two NFW halos \citep{Navarro1997} with virial masses $M_{200}=10^{11}$ and $10^{12} \rm M_\odot$ and concentration $c= 10$ and $8.3$ respectively following \citep{Dutton2014}. The dark gray shaded area indicate the changes expected when varying the concentration of the most massive NFW halo between $c=6.3$ and $c=10.3$, while light gray shows a more concentrated case with $c=20$ for illustration, substantially larger than what is expected for MW-mass halos today. 

In agreement with previous claims, most UDGs reported before this work are consistent with living in dwarf-mass halos with virial mass $\sim 10^{11} \rm M_\odot$, with perhaps the exception of DGSAT-1. Of the UDGs in this work, $4$ agree with such scenario and have mass estimates with errors overlapping well with the $10^{11} \; \rm M_\odot$ case or below. However, the majority of our sample seems to prefer larger mass halos, consistent not only with $10^{12}\; \rm M_\odot$ virial masses, but also overly concentrated in order to explain their large $M_{1/2}$ estimates within the half tracer radii. In other words: UDGs in our sample come in all flavors: from those consistent with no dark-matter, to those living in dwarf-mass halos up to the majority living in overly-concentrated MW-mass halos. 

In terms of their GC content, we find that the UDGs in our sample occupy the large-number end of the distribution at a given stellar mass, with expected numbers $10$-$30$ once corrected by incompleteness, radial distribution, and statistical contamination \citep[see][for details]{Lim2020}. In summary, the number of GC candidates in the NGVS images are counted within a $1.5R_e$ of each galaxy, after applying a statistical background correction, and then doubled to estimate the number of observable GCs within a galaxy. As this final number can still have contaminants, an additional correction is applied based on the numbers of GC candidates statistically found within local background regions. The final estimates for the total number of GCs are listed in Table~\ref{table:data}. 

As shown on the right panel of Fig.~\ref{fig:mass-NGC}, our UDGs show a modest excess of GCs compared to other galaxies. This bias is partially expected: we have selected all UDGs in the \citet{Lim2020} sample that have $10$ or more candidate GCs to increase the chances of obtaining accurate velocity dispersion measurements. We are therefore by default targeting objects with possibly the largest GC contents. 

Of course, other properties of GCs, besides their numbers, such as colors and luminosity functions may shed some light on the formation mechanisms of these galaxies. We defer such a detailed study to future work, not before highlighting that while our Virgo UDGs show an excess of GCs, only our most massive UDG reaches $\sim 100$ GCs, and hence comparable to the MW, while the rest show a more humble GC excess or are, in some cases, consistent with GC numbers seen in other early-type dwarfs. The GC content of these UDGs remains quite impressive: UDGs with the stellar content of the Fornax dSph show $20$-$30$ expected GCs, in contrast with the $\sim 5$ observed in Fornax dSph.

\section{Discussion}\label{sec:discussion}

In Section~\ref{sec:analysis} we show that UDGs have extreme kinematic and dynamical properties: (1) $50\%$ of our sample have velocity dispersions that are $\sim 8\times$ larger than those found in other quiescent galaxies of the same luminosity; (2) $80\%$ of our UDGs are extremely dark matter dominated ($f_{DM,1/2}>90\%$), (3) they have dynamical masses that are $\sim 100\times$  larger than that found in other galaxies with similar stellar mass; (4) $60\%$ of our sample have more GCs than expected in galaxies of the same stellar mass (either dwarf galaxies or the remaining Virgo UDGs not included in this study, see Section~\ref{sec:data}). 

\subsection{Extreme UDGs in Virgo} \label{sec:discussion_extreme}

The most important of our results is that a large number of UDGs in our sample (NGVSUDG-05, NGVSUDG-11, NGVSUDG-19, NGVSUDG-20 and NGVSUDG-A04) show very high velocity dispersion, $\sigma \sim 30 \rm - 80$ km/s, several times what is expected for dwarf galaxies of similar stellar mass. This suggests that they inhabit overly massive dark matter halos, providing tentative
 support to the original ``failed MW" or ``failed galaxy" scenario for the highest $M_{1/2}$ UDGs. In such context,  UDGs would inhabit massive dark matter halos that are more typical of galaxies $\sim 10$ to $100$ times brighter than themselves, requiring of additional physical processes to explain why the growth of the stellar component in the galaxy was truncated at an earlier stage compared to the growth of their dark matter halos. 

The processes involved in shutting down so dramatically their star formation are, however, not well understood.  One possibility is that the stellar growth truncation may be related to environmental processes such as ram pressure stripping or harassment in the cluster environments. However, no cosmological simulation has been able to naturally predict the formation of such extreme ``failed" systems, despite having the ability to model environmental processes that result in realistic galaxy populations. 

For instance, in current state-of-the-art cosmological numerical simulations such as TNG50, which models a cosmological volume large enough to include some spread in environment and accretion histories, UDGs are formed via high-spin halos and, to lesser degree, dynamical heating; with the expectation that they all inhabit dwarf-mass dark matter halos \citep{Benavides2022}. All other cosmological simulations that include high-density environments find a combination of processes involving feedback and environment, but in not a single case a failed-galaxy compatible with our measurement is predicted \citep{DiCintio2017,Chan2018,Jiang2019,Tremmel2020}. 

More exotic explanations for the formation of such systems have been proposed, for example invoking extreme conditions in the early universe leading to clumpy star formation with feedback so efficient that is able to self-quench the galaxy even in isolation \citep{Danieli22}. If such a scenario is plausible, it is yet to be predicted by our state-of-the-art cosmological models and most realistic feedback and star-formation treatments. The challenge is not minor: to generate objects as dark matter dominated as dSphs in the Local group, but in halos $10$-$100$ times more massive. It would also imply, in the case of our sample, that this special mode of star formation was quite common for UDGs in the environment such as Virgo. 

 These large velocity dispersions could, in principle, be also due to tidal stripping or out-of-equilibrium processes. However, \citet{Doppel2021} explore this possibility using GCs tagged onto the Illustris simulations, and find that, even for galaxies with small numbers of GCs, the true mass is recovered typically within a factor of two. These simulations contain nine different clusters with galaxies showing a wide range of dynamical states, including undergoing tidal disruption and any out-of-equilibrium processes that arises from their evolution within the galaxy cluster. Although there are some outliers for which inferred masses are significantly larger than the true value, it is unlikely that $50\%$ of our sample lies within these outliers. A detailed study exploring all of these possibilities will be presented in Doppel et al., in prep.

Our results are the first to kinematically suggest the possibility of UDGs as ``failed galaxies". This scenario has been invoked before for UDG candidates in low density regions, such as NGC~5846-UDG1 and DGSAT-I \citep[][respectively]{Danieli22, Janssens22}. However, the authors argue from different sets of observables related to the GC color and luminosities, not kinematics. In fact, DFX1 is consistent with the kinematics of the normal galaxies with the same luminosity in Figure~\ref{fig:sig_MV}, with only a modest mass-to-light ratio $\sim 30$ \citep{vanDokkum2017}. If this galaxy is indeed a ``failed galaxy", our sample of UDGs in Virgo show far more extreme level of ``failing" at forming their stars, as their mass-to-light ratios range from  $100$ to more than $1500~ \rm M_{\odot}/\rm L_{\odot}$. None of the seven UDGs from the literature with kinematics information are in the Virgo cluster. All of them are in isolation or very low density environments but for one, UDG7 \citep{Chilingarian2019}, that is in the Coma cluster but shows properties fully consistent with those of dwarf early-types (see Figures~\ref{fig:sig_MV} and \ref{fig:Mdyn-Mstar}). This reinforces the idea that such an extreme ``failing" scenario could be related to the dense environment of the Virgo cluster.

We have additional and complementary information for two of our UDGs. Our HST imaging for NGVSUDG-A04 and NGVSUDG-11 shows that these galaxies are at distances of $17.7^{+0.6}_{-0.4}$~Mpc~ and $12.7^{+1.3}_{-1.1}$~Mpc, respectively \citep[][Zhang et al. in prep.]{Mihos2022}. These distances suggest that, while NGVSUDG-A04 is on the far end of the cluster, NGVSUDG-11 is in front of the cluster most likely no longer gravitationally associated with it and, thus, now in isolation as there are no other foreground galaxies at that distance projected into the core of the Virgo cluster. This galaxy is therefore a good candidate to be a backsplash object, as found for simulated quenched UDGs in the field using the TNG50 simulations \citep{Benavides2021}. 


It is important, however, to consider the less exciting possibility that contamination by interlopers could be inflating the velocity dispersion estimates for our extreme UDGs. We are minimizing the presence of interlopers by selecting member GCs that are close in space and velocity to our UDGs (see Fig.~\ref{fig:membership}). Yet, given the low number of tracers used for the $\sigma$ determination, even the addition of a few contaminants could heavily impact the velocity dispersion measurement. The most likely contaminants are GCs in the intra-cluster component of the Virgo cluster or, in the case of UDGs with systematic velocity near $ \sim 0$ km/s, foreground stars in the MW. 

We have mitigated the impact of the foreground stars in the MW by restricting to GC candidates with the highest probability of being GCs based on their concentration index and colors (see Section~\ref{sec:membership}). This choice is further justified by our HST observations of NGVSUDG-11 and NGVSUDG-A04, which confirmed that all GC candidates with this highest probability are in fact resolved from space, confirming their GC nature. In addition, we use the Besan{\c{c}}on model to predict the expected number of MW stars that have the same apparent magnitude and colors as our selected GC candidates and are within the same solid angle and in the same line of sight of our selected sources. Using our selection function for each one of the Keck/DEIMOS masks, we find that for NGVSUDG-09, NGVSUDG-10, and NGVSUDG-19 no MW stars are expected within the GC satellite box defined in Figure~\ref{fig:membership}. In the case of NGVSUDG-11, there could be up to a maximum of two MW stars within our box of GC satellites. However, we have HST imaging for this galaxy showing that eleven of the GC satellites are resolved objects, confirming their GC nature, and including or removing the additional three, which are outside the field of view of the HST imaging, does not significantly change the velocity dispersion of this galaxy. Therefore, it is unlikely that foreground contamination significantly affects our results.

\begin{figure}[h]
\begin{center}
\includegraphics[clip=true,width=1.1\linewidth]{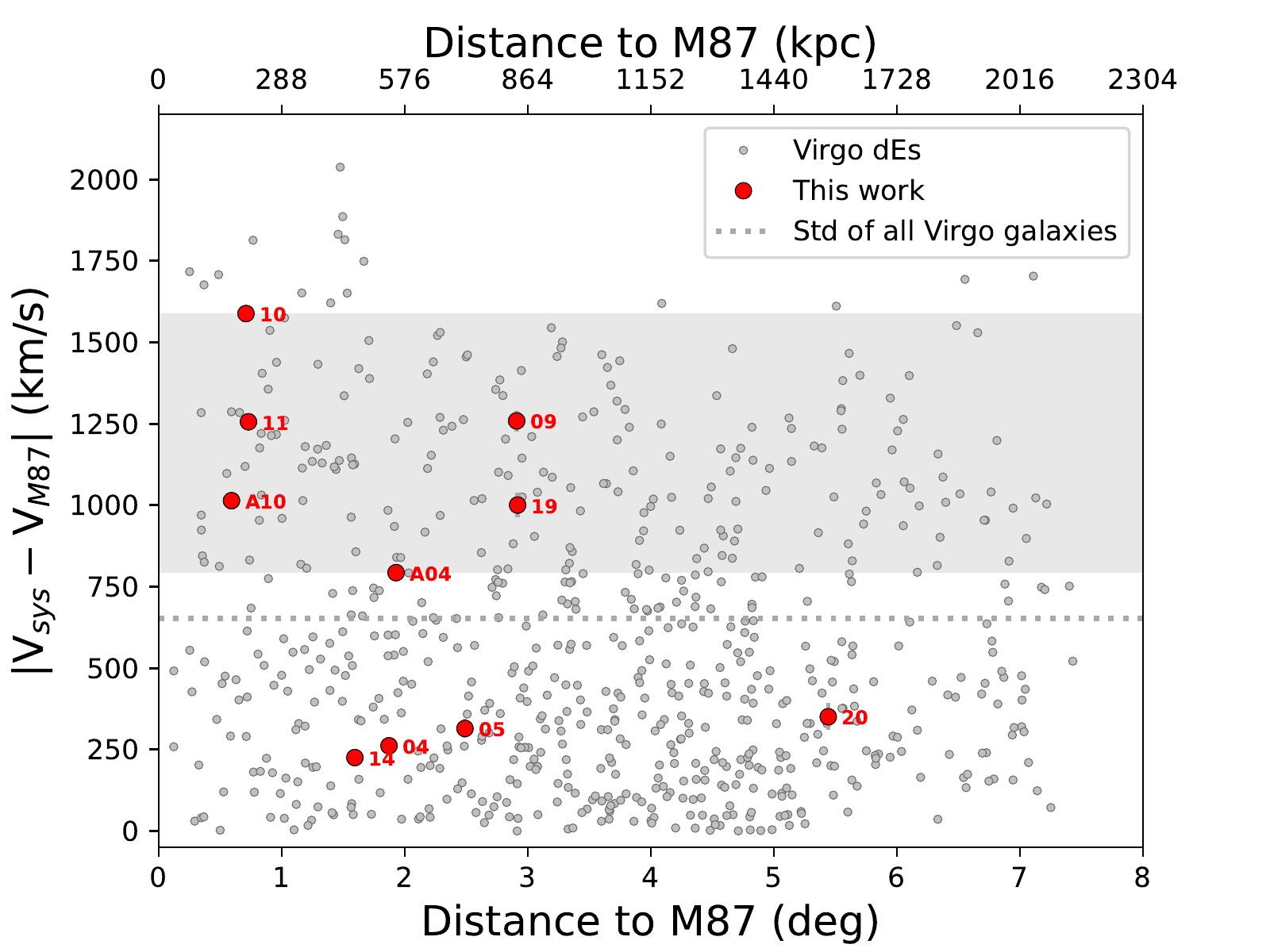}
\caption{Phase Space Diagram: Systemic velocity with respect to M87's as a function of projected distance to the center of the Virgo cluster, assumed to be at the location of M87. The UDGs presented in this work are shown in red. Gray circles indicate the distribution of dwarf early-type galaxies in the Virgo cluster that have the same luminosity range as our sample of UDGs. The dotted gray line shows the standard deviation of all Virgo galaxies, regardless their morphological classification or luminosity, with a systemic velocity measurement. Most of the UDGs show large relative cluster-centric velocities (highlighted by the gray shaded area), which makes the contamination by intra-cluster GCs less likely. Interestingly, UDGs are not clustered around the center of the Virgo cluster velocity as the rest of the Virgo members are. They seem to be a population with a large velocity dispersion within the cluster.} 
\label{fig:phase_space}
\end{center}
\end{figure}

\begin{figure}[h]
\begin{center}
\includegraphics[clip=true,width=1.1\linewidth]{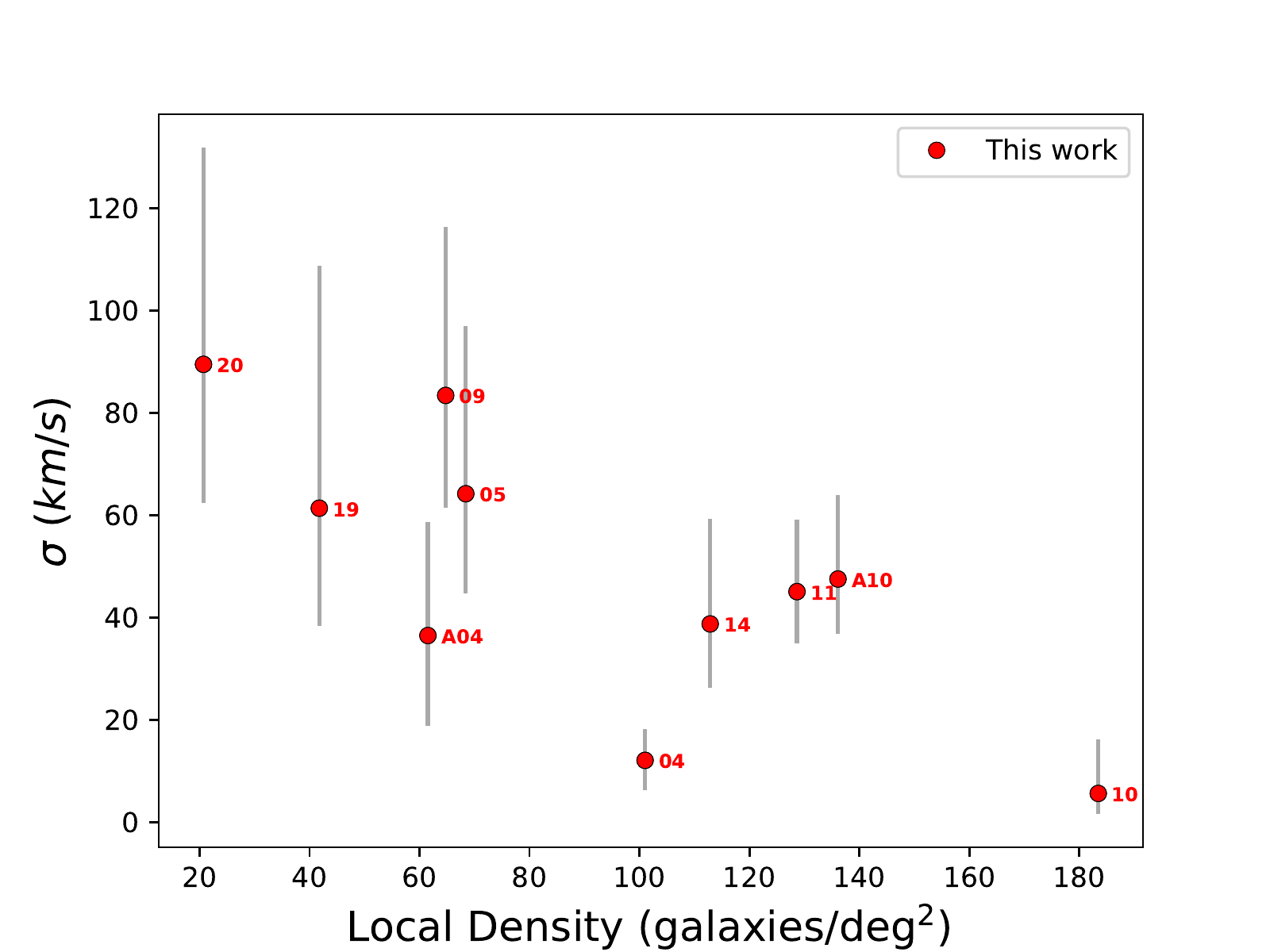}
\caption{Environment Diagram: Velocity dispersion with respect to the projected local density calculated as the distance to the tenth neighbor including all galaxies within the NGVS footprint. The outliers from the $\sigma$-luminosity relation (Figure~\ref{fig:sig_MV}) are all in the regions of low local density. NGVSUDG-11 is also quite isolated as it is in front of the Virgo cluster even though it is in projection close to M87, thus is high projected local density is not such when distance is taken into account. If only galaxies brighter than $M_g=-13.5$ are considered, all UDGs are in extremely low local density regions, more than $1.^{\circ}5$ from M87 and without any luminous galaxy close-by \citep[see Appendix~B by][for notes on individual UDGs]{Lim2020}. The only galaxy with somewhat more luminous dwarf galaxies close is NGVSUDG-10, with three VCC galaxies within $13'$. However, these three galaxies (VCC1047, VCC1036, and VCC1010) have all velocities close to $\sim 1000$~\kms~ (724~\kms, 1124~\kms, 934~\kms, respectively), while NGVSUDG-10 has a systemic velocity of $-292$~\kms.} 
\label{fig:localdensity}
\end{center}
\end{figure}

Contamination by intra-cluster GCs, on the other hand, is in principle possible for UDGs with velocities more typical of the Virgo mean velocity of $\sim1000$~km~s$^{-1}$. Encouragingly, a point to argue against significant contamination from intra-cluster GCs in Virgo is the fact that several of the extreme UDGs have also quite large cluster-centric relative radial velocities ($|V_{sys}-V_{M87}|>750$)~km~s$^{-1}$ (see Figure~\ref{fig:phase_space}), which would make the likelihood of overlap in velocity-space between their associated GCs and the intra-cluster GCs of Virgo quite small. Figure~\ref{fig:localdensity} shows that the UDGs with the highest $\sigma$ are in fact located in regions of low projected local density, which means that they are surrounded by a small number of neighbors. The vast majority of those neighbors are dwarf galaxies with $M_g>-13.5$, such low luminosity dwarf galaxies rarely have any GCs and, if they do, they do not extend spatially far enough to be within our selection box \citep{Peng2006}. A special case of intra-cluster GCs contamination is NGVSUDG-10 and NGVSUDG-11, the only galaxies in our sample that are closer both in projected distance and velocity to M86, an early-type galaxy member of the Virgo cluster with a systemic velocity of $\sim -220$~\kms \citep{Cappellari2011}, than to M87. In this case, resolved objects with negative or close to zero velocities could be M86 GCs. However, NGVSUDG-10 has such a low $\sigma$ that its dark matter is negligible. If there are some M86 GCs contaminating our sample, they are not significantly increasing the dispersion of this galaxy. In the case of NGVSUDG-11, removing one single random GC from our sample does not change the measured velocity dispersion (see details in Section~\ref{sec:MCMC}). To reconcile the current position of NGVSUDG-11 in the $M_V-\sigma$ relation, its dispersion needs to be $\sim 30$~\kms~ smaller. Even by nearly quadrupling the sample of GCs presented in \citet{Toloba2018} for this galaxy, its $\sigma$ remains the same. This gives us confidence that, if some contaminants are present, $\sigma$ is not dramatically affected. 

NGVSUDG-09, NGVSUDG-19, and NGVSUDG-A04 belong to the high-velocity tail in phase space, are also large $\sigma$ objects, and are quite isolated without any nearby bright galaxy. Since the velocity distribution of intra-cluster GCs in Virgo is expected to be more consistent with the virialized, more bound regions of the cluster, the contamination in these extreme UDGs is n\"aively expected to be small. Late-type galaxies have a broader velocity distribution within the cluster \citep[e.g.][]{Boselli2006}, however, their dispersion of $\sim 820$~km~s$^{-1}$ is still not enough to explain the more extreme velocity distribution found in this sample of UDGs, where the majority of the UDGs are beyond $\sim 800$~\kms~ from the center of the velocity distribution. This suggests that UDGs with large amounts of GCs have an extreme velocity distribution within the Virgo cluster, possibly even larger than that of late-type galaxies. A more detailed evaluation of contamination effects expected for these UDGs will be presented in upcoming work (Doppel et al., in-prep). Confirmation and validation of the ``failed galaxy" scenario for the candidates presented here will ultimately require significant observing time commitment to measure velocity dispersions from the stellar component in the body of these galaxies.     

An additional consideration is whether these systems are in equilibrium. Small deviations from equilibrium do not significantly affect the velocity dispersion measured \citep[see][where they analyze how skewness and kurtosis of the Gaussian distribution of tracers do not significantly affect the velocity dispersion measured]{Doppel2021}. 

\subsection{Less extreme UDGs in Virgo}

Considering our full sample, the results support the scenario where more than one mechanism of formation is necessary to explain the properties of UDGs, even if a restrictive selection criteria, like the one assumed here following \citet{Lim2020}, is applied. We discuss some of the less-extreme cases individually below and interpret them in view of the formation paths discussed in \citet{Sales2020}. Using $14$ Virgo-like clusters from the Illustris TNG100 cosmological hydrodynamical simulation \citet{Sales2020} finds two main paths to form cluster UDGs: (a) born-UDGs are field low surface brightness galaxies that recently ($\sim 6$~Gyr ago median, albeit with significant dispersion) entered the cluster and now show large amounts of dark matter and have stellar populations with intermediate ages and low metallicities; (b) tidal-UDGs are higher surface brightness galaxies that fell into the cluster early-on and, as a consequence, have little to no dark matter left, old stellar ages and high metallicities. Note that born-UDG could correspond to several internal mechanisms discussed in the literature, such as strong feedback \citep{DiCintio2017,Chan2018} or high-spin halos \citep{AmoriscoLoeb2016,Benavides2022}. 


Five of the Virgo UDGs studied here may not be as extreme as the rest of our sample. NGVSUDG-10 lacks dark matter, i.e. its velocity dispersion is explained by stars alone. NGVSUDG-04 has an extremely low velocity dispersion, consistent with dSphs of the same luminosity. These are the only two galaxies with morphological hints of tidal tails, which could explain their lower dark matter content. In Appendix~\ref{A1} we investigate the possibility of these galaxies showing some rotation. However, the results are not conclusive and are consistent with no velocity gradients. These two galaxies could be consistent with being tidal-UDGs.
NGVSUDG-A10, NGVSUDG-14, and NGVSUDG-09 have consistent velocity dispersions with dwarfs of the same stellar mass, albeit an elevated dark matter fraction which might be explained by their more extended sizes compared to dEs, which could be consistent with a born-UDG scenario.

Having little to no dark matter is expected for low mass galaxies (dEs) in the Virgo cluster \citep{Toloba2014b}, however, the stellar mass range of these UDGs has not been dynamically studied in the Virgo cluster due to the extremely low surface brightness of these counterpart dEs. 
Studies targeting the properties of dEs with $M_* \lesssim 10^8 \; \rm M_\odot$ in Virgo are necessary to confirm the hypothesis that these objects are consistent with the extended tail of normal dwarfs with the same stellar mass (low mass dEs) or ``born-UDGs", or if they are better explained by more massive progenitors that got heavily tidally stripped ``tidal-UDGs". Information about their stellar populations would help to answer this question, as massive early-type galaxies in any environment are on average older and more metal rich than Virgo cluster dEs \citep{Toloba2014b,McDermid2015}.

\section{Summary}\label{sec:summary}

We present kinematics ($V_{sys}$ and $\sigma$) and derived dynamical masses for a sample of ten UDGs found in the Virgo cluster using spectroscopic data of their associated GCs from Keck/DEIMOS. This is the most numerous, uniform and most strictly selected sample of UDGs with kinematical information to date. The sample was homogeneously selected from a full analysis of the structural parameters of all galaxies detected in the NGVS footprint \citep[][]{Ferrarese2012}. All of our UDGs were selected to be more than $2.5\sigma$ outliers in at least one of the scaling relations that combine luminosity, size, and surface brightness \citep[see ][]{Lim2020}. Our final sample of ten Virgo UDGs contains all UDGs with more than $10$ GC candidates, $8$ of them are outliers in all three scaling relations while $2$ are outliers in at least one.  

MCMC analysis of the kinematics of these UDGs obtained from their spectroscopically confirmed GC satellites reveals a surprising scatter in their velocity dispersion, with about half of our sample following the expected $\sigma-M_V$  relation defined by of other early-type galaxies and the other half showing substantially higher velocity dispersions, $\sigma \sim 30$ - $90$~\kms, which is between $\sim 3$-$8$ times what is expected for these galaxies given their luminosity. The diversity in the kinematics of this sample includes also a case consistent with very little or ``dark-matter free" galaxy, NGVSUDG-10, which in this case may be explained by an ongoing case of tidal disruption.

The majority, $60\%$, of the sample has extreme systemic velocities. While the center of Virgo is at $\sim 1200$~\kms~, most of our galaxies have velocities below $\sim 300$~\kms~ and above $\sim 2000$~\kms. In addition, the UDGs with the highest $\sigma$ are in low density regions of the cluster, with only a few extremely low luminosity dwarf galaxies close by. These findings may suggest that UDGs, or at least those with large numbers of GCs, are a hot galaxy component within the Virgo cluster, with a dispersion larger than that of spiral galaxies. Most likely, the environment is playing an important role. 

UDGs in the high-$\sigma$ subsample provide, for the first time, kinematical support to the scenario that at least some UDGs might be ``failed  galaxies". Previous studies used indirect diagnostics to propose that at least some UDGs could be hosted by overly massive dark matter halos, including arguments based on their GCs number, GC luminosity functions or colors \citep{vanDokkum15,Peng2016, Lim2018, Lim2020, Danieli22, Janssens22}. With the possible exception of DGSAT-1, all previous UDGs in the literature  with kinematical data are instead consistent with normal dwarf halos and all of them, but for one, are found in low density environments. Our results are potentially an important paradigm shift from the settling scenario where all UDGs are the most extended tail of the dwarf galaxy population, since the kinematics and high mass-to-light ratios inferred for our high-$\sigma$ sample are only compatible with massive halos, comparable or above that of the MW. 

There are, however, caveats to be considered before rushing to conclusions. The most important of them is the possible role played by interlopers or contaminants in our identified population of associated GCs. Many of our UDGs have more or close to 10 GCs labeled as gravitationally bound and therefore participating of the velocity dispersion measurement. While numerical simulations suggest that such number is enough to provide an unbiased measurement of intrinsic kinematics \citep{Toloba2018, Doppel2021}, and that half of that tracers still provides nearly negligible offsets, it is still a relatively low number that may be biased if even a few interlopers are included in the calculation. 

Expected contaminants might come mostly from two sources: foreground stars in the MW for Virgo objects with $\sim 0$ systemic velocity or intra-cluster GCs free-floating in the gravitational potential of the Virgo cluster. By using the concentration index and colors method informed with HST data we believe to have mitigated most of the impact from foreground stars in the MW (see Sec.~\ref{sec:analysis}). In addition, estimations using the Besan{\c{c}}on model \citep{BesanconModel} and our selection function indicate that less than one MW star may have ended up within the GC satellite selection box. The fact that several of our potential ``failed MW" UDGs also show extreme cluster-centric radial velocities and that all of them are in quite isolated areas of the Virgo cluster only surrounded by extremely low luminosity dwarf galaxies makes it less likely to be explained by contamination from intra-cluster GCs in Virgo, however this cannot be fully be ruled out without further data.

Regarding the total number of GCs in our calculations, in \citet{Toloba2018}, with only 4 GCs we estimated  $\sigma=47^{+53}_{-29}$~\kms~ for NGVSUDG-11; in this work, with a total of 14 GCs, the new dispersion is $\sigma=45^{+14}_{-10}$~\kms. More than three times the number of GCs and still the velocity dispersion did not change much. This may not happen for all of our galaxies, but real data and simulations like those in \citet{Doppel2021}, suggest that the number of GCs considered here is enough to estimate dispersions that should not exhibit offsets much larger than 5-10~\kms, which are extremely small offsets in comparison to the extreme velocity dispersions measured for their luminosities.

Taken at face value, our results contribute two important points to our understanding of UDGs. First, there must be more than one mechanism to form such a diverse population. For those consistent with dwarf-mass halos, models that combine ``born" UDGs with ``tidal" UDGs seem to be the most adequate \citep{Jiang2019,Sales2020}. Second, for the puzzling population consistent with the ``failing galaxy" scenario, there is yet much to be understood, as the physical mechanisms leading to their dramatic stellar growth truncation within such inferred massive dark matter halos are currently missing in all our state-of-the-art cosmological simulations of galaxy formation. Moreover, such ''failed galaxies" mean a potentially serious revision to the idea that galaxy formation efficiency is, to first order, determined solely by halo mass.

Our UDG sample selected to have a large number of GCs shows at least 50\% of them belonging to this high velocity-dispersion category, suggesting that whatever the process is turning these halos extremely inefficient to form stars, it should be relatively common in high-density regions such as galaxy clusters. Follow up spectroscopic studies of the stellar components in these galaxies are necessary to confirm the velocity dispersion measurements presented here without uncertainties associated to GC membership as well as to constrain their star formation histories, possibly revealing clues on the reason for their ``failure" to grow galaxies $\sim 10$-$100$ times brighter as they were destined to form according to their halo kinematics.  

\begin{acknowledgments}
The authors would like to thank insightful discussions with Jessica Doppel, Anna Ferre-Mateu, Julio Navarro, Aaron Romanowsky, and Guillermo Barro. {bf In addition, the authors would like to thank the anonymous referee for making suggestions that helped improve this manuscript.}
ET is thankful for the support from HST GO program GO-15417 and from NSF-AST-2206498 grants. LVS is grateful for partial financial support from NASA-ATP-80NSSC20K0566 and NSF-CAREER-1945310 grants. SL acknowledges the support from the Sejong Science Fellowship Program through the National Research Foundation of Korea (NRF-2021R1C1C2006790). The spectroscopic data presented herein were obtained at the W.M. Keck Observatory, which is operated as a scientific partnership among the California Institute of Technology, the University of California and the National Aeronautics and Space Administration. The Observatory was made possible by the generous financial support of the W.M. Keck Foundation. The authors wish to recognize and acknowledge the very significant cultural role and reverence that the summit of Maunakea has always had within the indigenous Hawaiian community. We are most fortunate to have the opportunity to conduct observations from this mountain.
\end{acknowledgments}

%

\vspace{5mm}
\facilities{Keck(DEIMOS), HST(WFC3), CFHT(MegaCam)}



\software{astropy \citep{2013A&A...558A..33A,2018AJ....156..123A}
          }



\appendix

\section{Velocity Gradient}\label{A1}

We investigate whether the GCs in these UDGs show any velocity gradient. For this task, we use the MCMC implementation described in \citet{Toloba2018}. The logarithmic probability that includes a velocity gradient has the following shape:

\begin{equation}\label{eq:MCMC_vgrad}
\mathcal{L} (V_{sys}, \sigma, dv/dr, \phi) = -\frac{1}{2}\sum_{n=1}^N \log (2\pi (\sigma^2+\delta
v_n^2)) - \sum_{n=1}^N
\frac{(v_n-V_{sys}-\frac{dv}{dr}r_n)^2}{2(\sigma^2+\delta v_n^2)}
\end{equation}

\noindent where $dv/dr$ is the velocity gradient along the projected distance $r$ with position angle $\phi$, which is calculated as follows:

\begin{equation}
r=(RA-RA_0)\cos (Dec_0) \sin (\phi) +(Dec-Dec_0)\cos (\phi)
\end{equation}

\noindent $RA_0$ and $Dec_0$ are the coordinates of the photometric galaxy center.

None of our galaxies have enough GCs to run this MCMC implementation with four free parameters ($V_{sys}$, $\sigma$, $dv/dr$, $\phi$). Even when reducing it to three free parameters by assuming that $\phi$ coincides with the semimajor axis of the galaxy, there is no clear evidence of a velocity gradient in any of the galaxies. All priors used in this MCMC implementation are flat priors within plausible physical ranges: $-500 < V_{sys} < 3000$~\kms; $0 < \sigma < 200$~\kms; and $-30 < dv/dr <30$~\kms~arcmin$^{-1}$.

\begin{figure*}[h]
\begin{center}
\includegraphics[clip=true,width=0.39\linewidth]{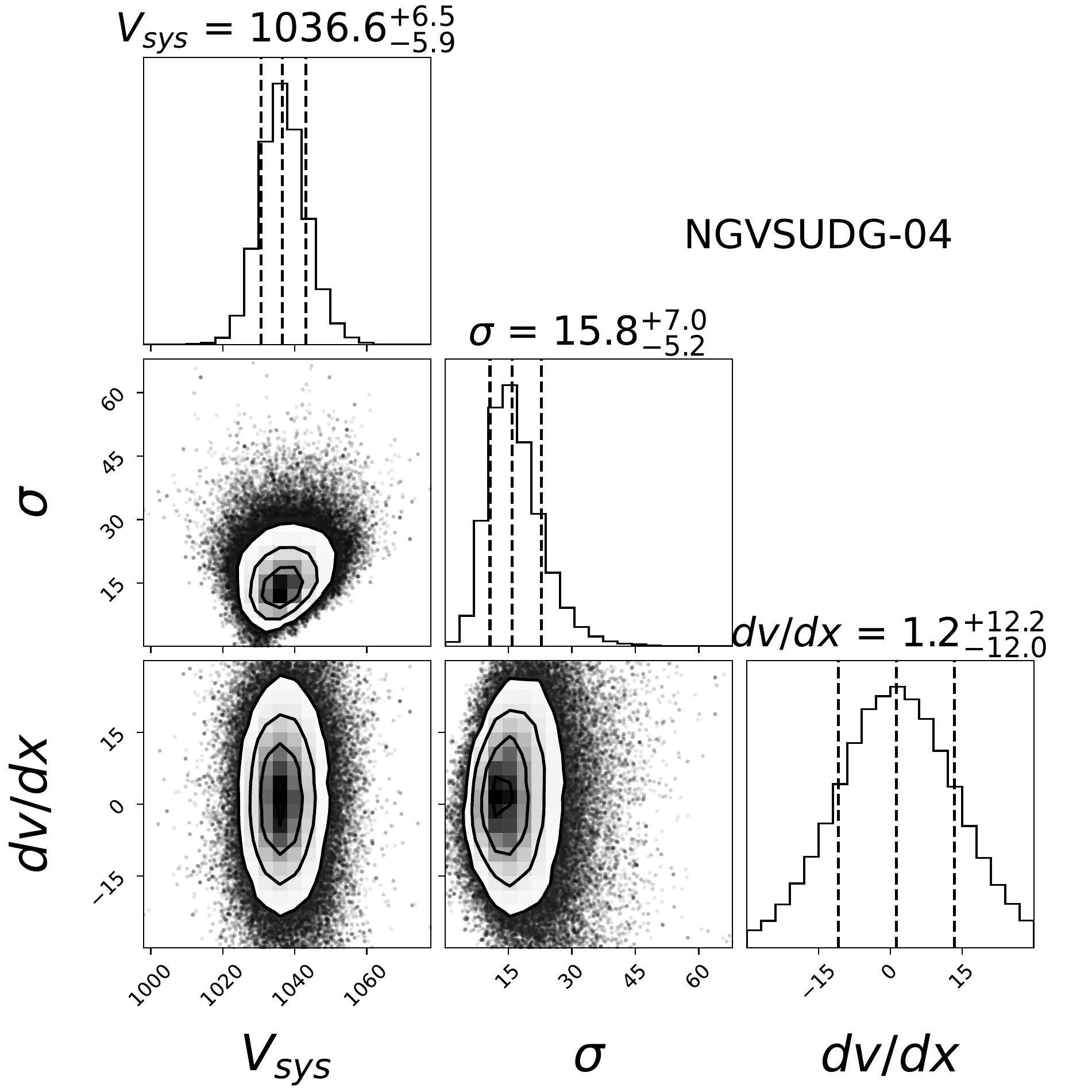}
\includegraphics[clip=true,width=0.39\linewidth]{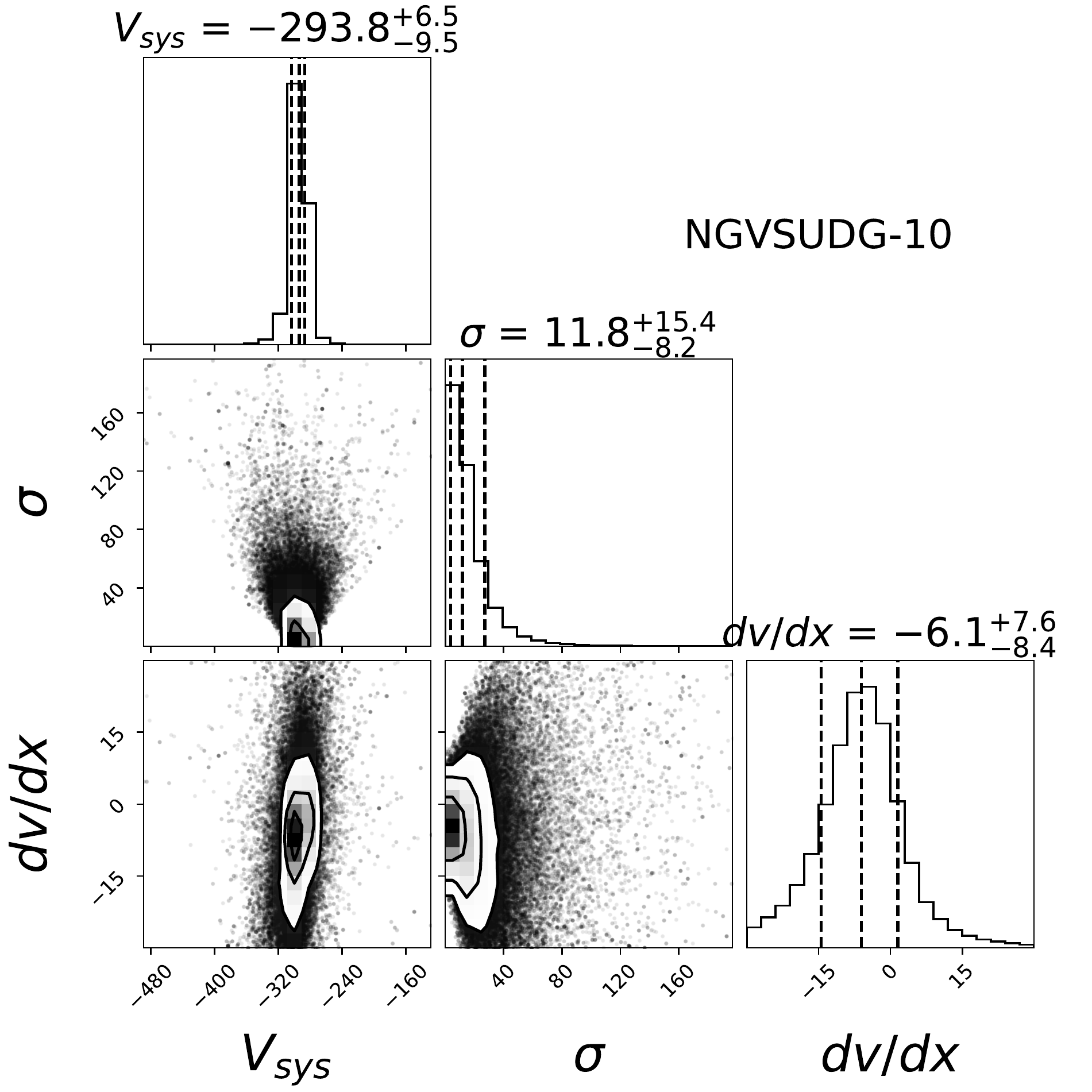}
\caption{Corner plots for NGVSUDG-04 ({\it left}) and NGVSUDG-10 ({\it right}). These plots show the two-dimensional and marginalized posterior probability density function for the systemic velocity $V_{sys}$, velocity dispersion $\sigma$, and velocity gradient $dv/dr$ estimated following the likelihood probability described in Equation~\ref{eq:MCMC_vgrad} when the position angle $\phi$ is fixed to the semimajor axis of the galaxy. Only for these two galaxies, the MCMC converges to a value. Due to the small statistics, the resultant velocity gradients have large uncertainties that makes them consistent with no gradient. Note that the $V_{sys}$ and $\sigma$ do not coincide with the values reported in Figure~\ref{fig:cornerplot} due to a different MCMC implementation (three free parameters instead of two) and the flat priors used here instead of the Jeffreys prior used in Figure~\ref{fig:cornerplot}.}

\label{figA:corner}
\end{center}
\end{figure*}

Figure~\ref{figA:corner} shows the MCMC results for NGVSUDG-04 and NGVSUDG-10, the only two for which the MCMC shows proper convergence, however, the velocity gradients estimated show large uncertaintites and are consistent with zero. The values obtained for NGVSUDG-04 agree within the error bars with our previous results \citep{Toloba2018}. These are also the only two galaxies in our sample that show some stellar elongation that may be indicative of tidal interactions.

\bibliography{refs}{}
\bibliographystyle{aasjournal}



\end{document}